\documentclass{aa}
\usepackage{graphicx}
\usepackage{txfonts}
\usepackage{hyperref}
\usepackage[normalem]{ulem}
\hypersetup{colorlinks = true, citecolor = {blue}, urlcolor= {blue}}

\newcommand{\msun}{{\rm M}_{\odot}}
\newcommand{\lsun}{{\rm L}_{\odot}}
\newcommand{\rsun}{{\rm R}_{\odot}}

\newcommand{\eg}{e.g.,\@\xspace}

\def\gapprox{\;\rlap{\lower 3.0pt                       
        \hbox{$\sim$}}\raise 2.5pt\hbox{$>$}\;}
\def\lapprox{\;\rlap{\lower 3.1pt                       
        \hbox{$\sim$}}\raise 2.7pt\hbox{$<$}\;}

\newcommand{\be}{ \begin{equation} }
\newcommand{\ee}{\end{equation}}

\newcommand{\ben}{\begin{enumerate}}
\newcommand{\een}{\end{enumerate}}

\renewenvironment{thebibliography}[1]{\begin{oldthebibliography}{#1}\setlength{\parskip}{0ex}\setlength{\itemsep}{0ex}}{\end{oldthebibliography}}

\usepackage{threeparttable}
\usepackage{diagbox}
\usepackage{siunitx}
\usepackage{ltablex}
\newcommand{\orcid}[1]{\href{https://orcid.org/#1}{\protect\includegraphics[width=8pt]{./figures/orcid.pdf}}}
\makeatletter
\renewcommand*\aa@pageof{, page \thepage{} of \pageref*{LastPage}}
\makeatother

\usepackage[dvipsnames]{xcolor}

\definecolor{darkgreen}{RGB}{31, 207, 31}

\usepackage{supertabular}

\usepackage{booktabs}

\newcommand{\ceph}{Cephe{\"\i}d}
\newcommand{\cephs}{Cephe{\"\i}ds}

\begin{document}

    \title{Implications of a turbulent convection model for classical Cepheids}

    \author{M. Deka
          \inst{1,2,3},
          F. Ahlborn
          \inst{4},
          T. A. M. Braun
          \inst{3,5},
          \and
          A. Weiss
          \inst{3}
          }

   \institute{INAF-Osservatorio astronomico di Capodimonte, Via Moiariello 16, I-80131 Napoli, Italy\\
              \email{mami.deka@inaf.it}
        \and
              Department of Physics, Cotton University, Panbazar, Guwahati 781001, Assam, India
         \and
             Max-Planck-Institut für Astrophysik, Karl-Schwarzschild-Straße 1, 85748 Garching, Germany
        \and
             Heidelberg Institute for Theoretical Studies (HITS), Schloss-Wolfsbrunnen Weg 35, Heidelberg, Germany 
        \and 
            Ludwig-Maximilians-Universität München, Geschwister-Scholl-Platz 1, 80539 Munich, Germany
             }

   \date{}

\abstract
   {
   The appearance of blue loops in the evolutionary tracks of intermediate-mass core He-burning stars is essential for explaining the observed characteristics of \cephs. The blue loops for lower mass \cephs\ cannot always be reproduced when only classical, local mixing length theory (MLT) is used. Additionally, classical models result in a mass discrepancy compared to pulsational and dynamical mass determinations. Both problems can be resolved through an ad-hoc extension of the MLT for convection. 
   }
   {We use the non-local Kuhfuss turbulent convection model (TCM) which can explain overshooting
   directly from the solution of the TCM equations. The primary objective of this study is
   to test the predictions of the Kuhfuss TCM when applied to intermediate-mass core He-burning stars and validate the model predictions against observations of \cephs.}
   {We used the state-of-the-art 1D stellar evolution code GARSTEC with the implementation of the Kuhfuss TCM and computed evolutionary tracks for intermediate-mass core He-burning stars. We compare these tracks with those computed with MLT including and excluding ad-hoc overshooting and with observations of five \cephs\ in detached binary systems obtained from the literature.} 
   {The stellar evolution tracks generated using the Kuhfuss TCM and MLT with ad-hoc overshooting exhibit similar appearances. Overshoot mixing from the convective boundaries and the occurrence of the \ceph\ blue-loop have been achieved naturally as solutions to the equations of the Kuhfuss TCM. Furthermore, the evolutionary models including the Kuhfuss TCM have been successful in reproducing the observed stellar parameters, including mass, luminosity, radius and effective temperature.  
   }
   {We have successfully generated \cephs' blue loops with a TCM without any fine-tuning of the involved numerical 
   parameters and with overshooting predicted directly from the convection theory. Beyond the achievement of blue loops, our approach which treats convection more physically has also been able to reproduce the observations of \cephs\ in eclipsing binary systems with a similar accuracy as MLT models with ad-hoc overshooting.
   }

   \titlerunning{Turbulent convection model for stellar evolution}
   \authorrunning{M.\ Deka et al.}

   \maketitle
   
\section{Introduction}
\label{sec:Introduction}
Convection is one of the most important mechanisms for the transport of energy, angular momentum and chemical elements within stars.
It significantly impacts the stellar structure, evolution and pulsation properties. However, convection within stars is a highly turbulent, non-linear and time-dependent three-dimensional (3D) phenomenon, resulting in considerable 
computational demands. Furthermore, the implementation of 3D turbulent convection into simplified one-dimensional (1D) calculations is very challenging. 

In the literature, the prevailing approach for describing convection is the widely adopted Mixing Length Theory (MLT) 
\citep{bier32,bohm58}. However, MLT is time-independent and local in nature. It considers the boundary of convectively unstable regions (the so-called Schwarzschild or Ledoux boundary) as rigid, thus falling short in explaining convective motions beyond this boundary. This is unphysical, because when convective fluid parcels approach the boundary with non-zero velocity, the velocity cannot drop to zero immediately at the Schwarzschild boundary, and the fluid parcel will penetrate some distance into the stable layers to decelerate. Moreover,
convective eddies are expected to rise above the Schwarzschild boundary from 3D simulations as well as observations \citep[\eg][]{ande91,pate93,capu00,piet04,kell08,andr15,ande22,higg23}. 
For example, the standard MLT models without further additions do not produce blue loops extended enough in the evolutionary tracks of classical \cephs\ (hereafter \cephs\ for short) for certain masses. \cephs\ are a class of intermediate-mass pulsating stars for which the existence and extent of the blue loops are essential. To account for convective overshooting effects in stellar models, the standard MLT can be modified. The influence of convective eddies rising into the formally stable layers above the Schwarzschild boundary can be mimicked by introducing additional mixing beyond this boundary—a concept we refer to as ``ad-hoc overshooting''.  
%
This is done either by extending the chemical 
mixing beyond the Schwarzschild boundary by a certain fraction of the pressure scale height (step overshooting) or by using a diffusive process \citep{frey96}, where the diffusion coefficient decreases exponentially with the distance from the Schwarzschild boundary (exponential or diffusive overshooting).

The inclusion of ad-hoc overshooting in MLT has successfully produced 
the \ceph\  blue loops \citep{capu00,kell08,cass11,mill20,bono24}. The parameters of this approach can be calibrated to observations. However, there is no physical justification to generalize these parameter values, calibrated to a certain set of observations, to all possible applications. Besides, it fails to predict the temperature gradient within the overshooting region or the extent of this region. Furthermore,
due to its time-independent nature, the MLT models are not suitable for interpreting any dynamical process whose characteristic time scale is comparable to that of convection. One such crucial process is stellar pulsation. This underlines the need to incorporate more physically grounded, numerically feasible and non-local theories of convection such as a ``turbulent convection model'' (TCM) into stellar structure and evolutionary/pulsation models. TCMs are derived from the basic hydrodynamic conservation equations, which describe the dynamics of the turbulent convective motions. Convective overshooting naturally emerges from solving these equations without requiring any ad-hoc descriptions. Additionally, TCMs provide the convective flux, enabling the prediction of the temperature gradient in the overshooting region.

A large number of TCMs developed for stellar convection can be found in the literature \citep{xion78,xion86,stel82,kuhf86,kuhf87,canu92,canu93,canu97,canu98,canu11,li07}. These models vary in terms of the chosen variables, approximations, and assumptions \citep[][and references therein]{kupk22}. The TCM developed by \citet{kuhf86,kuhf87} exists in two versions: the 1-equation \citep{kuhf86} and 3-equation \citep{kuhf87} model. The original version of the Kuhfuss TCM has undergone improvements by \citet{wuch98}, \citet{flas03}, \citet{kupk22}, and \citet{ahlb22}.  In the present study, we use the improved version of 
the Kuhfuss 1-equation TCM, which has already been implemented into the Garching Stellar Evolution Code \citep[GARSTEC;][]{weis08} \citep[for details see][]{flas03,kupk22,ahlb22,tere24}. 
This model has already undergone thorough testing in the context of convective cores of intermediate-mass main-sequence stars and has demonstrated good agreement with models using MLT with ad hoc overshooting, which were tuned to match observations \citep{ahlb22}. However, the prediction of a standard solar model calculated with the Kuhfuss 1-equation TCM does not align well with 
the observed sound-speed profile and the depth of the convective envelope of the Sun
\citep{tere24}. This discrepancy necessitates further testing of the Kuhfuss TCM in stellar models across different mass ranges and different evolutionary stages.

In the present study, we focus on testing and extending the mixing properties of the 1-equation model and how they perform when applied to intermediate-mass core He-burning stars, particularly \cephs. \cephs\  represent a class of intermediate-mass pulsating stars in the core He-burning phase. They are particularly known because of their period-luminosity relation (PLR), which makes them excellent distance indicators \citep{leav08,leav12,bhar23}. Besides being crucial 
distance indicators, \cephs\  also provide us with ideal stellar laboratories to test and validate stellar evolutionary and pulsation theories because of the additional information we can obtain about these stars due to their pulsations.

In order to test the 1-equation model, we make use of the ``\ceph\  mass discrepancy'' problem \citep{stob69,cox80,kell08} which is known to be sensitive to the treatment of convection in the evolutionary models \citep{cass11}.
The ``\ceph\  mass discrepancy'' problem is a long-standing puzzle in the field of stellar evolutionary and pulsation theories.
Nevertheless, this has presented us with a unique opportunity to refine theoretical frameworks. A \ceph's mass can 
be determined using both its evolutionary and pulsation properties. However, the \ceph's evolutionary mass has been 
found to be significantly higher than the pulsation mass \citep{stel82}. \citet{cox80} reported this mass discrepancy to be 
$\sim40\%$. The improved opacity calculations performed by the OPAL group \citep{igle93,igle96} have reduced this discrepancy
to $\sim(10-20)\%$. Nevertheless, it was not clear whether the evolutionary mass or pulsation mass was correct until recently. 

\citet{piet10} for the first time measured the dynamical mass of \cephs\  to an accuracy of $1\%$ for the OGLE-LMC-CEP-0227 
double line eclipsing binary system. Their findings indicate consistency between the dynamical mass and the pulsation mass within the margins of error.
Hence, the possible solutions for the remaining discrepancy concern the improvement of the physics of the evolutionary model such as convective core overshooting, mass loss, and rotational mixing \citep[][and references therein]{kell08,neil11,moro12,ande14}. Convective core overshooting during main sequence (MS) evolution leads to a more massive post-MS helium core, thereby resulting in a more luminous \ceph\  for the same mass \citep{cass11}. Incorporating mass loss during the \ceph\  evolution can reduce the stellar mass without affecting the stellar luminosity significantly \citep{neil11}. Finally, including
rotation can also increase the luminosity of the \ceph\  without increasing the initial mass 
\citep{ande14}. The model fitting of light curves and radial velocity and radius time series of the \ceph\ by \citet{marc13} also provided  an indication for a slightly non-canonical mass-luminosity relation to be in better agreement with observation. Thus, including these effects in evolutionary models will help us to bring the evolutionary mass and the dynamical mass into agreement. The present study focuses on the more physical treatment of convective core overshooting. Our goal is to test the TCM by \citet{kuhf86} by comparing model predictions with observations of \cephs\ in binary systems.
In all cases the mass of the \ceph\ is close to $4\,M_\odot$, and thus on the lower side of the \ceph\ mass range, where classical MLT models
either do not produce blue loops or produce blue loops that do not cross the Cepheid instability strip  (cf.\ 
Fig.~\ref{fig:all_hrd_5M}).

The remainder of the paper is organized as follows: we give a brief mathematical overview of the Kuhfuss 1-equation model (TCM1) in Sect.~\ref{sec:2.1}.  In Sect.~\ref{sec:ev_models}, we discuss the computation of stellar models with TCM1 and the `` ad-hoc MLT plus overshooting'' approach, and the different parameters adopted
in the calculations. Then, we compare TCM1 results with the MLT model without and with the addition of overshooting in Sect.~\ref{sec:3.1}. 
In Sect.~\ref{sec:3.2}, we explore how applying non-locality to different convective boundaries in a star affects the blue loops.
Sect.~\ref{sec:3.3} presents the models for the five \ceph\ binary systems we studied (OGLE-LMC-CEP-0227, OGLE-LMC-CEP-1812, OGLE-LMC-CEP-4506, OGLE-LMC-CEP-2532 and OGLE-LMC-CEP-1718). We discuss the impact of the various uncertainties involved in both theory and observations in Sect.~\ref{sec:discussion}. Finally, the conclusions and future prospects of our findings are presented in Sect.~\ref{sec:conclusion}. 

\section{Stellar evolutionary models with TCM}
\label{sec:method1}
\subsection{The Kuhfuss 1-equation turbulent convection model}
\label{sec:2.1}
\citet{kuhf86,kuhf87} developed a TCM by applying the Reynolds stress formalism to the fundamental hydrodynamic equations. The TCM incorporates non-locality and the time-dependence of convection. 
This allows the TCM to quantify the extent of the convective boundary mixing (CBM) region based on non-zero velocities as well as the temperature gradient within this region.

The TCM of Kuhfuss exists in two versions: 1-equation and 3-equation model \citep{kuhf86,kuhf87}. The 1-equation version (hereafter TCM1) is the simplified version that determines the turbulent kinetic energy (TKE; $\omega$) using a single equation
including the non-local terms. In the stationary and local limit, the standard MLT equations can be derived from this model. 
In the 3-equation formalism (hereafter TCM3), two additional equations describing the convective flux ($\Pi$) and the second order entropy fluctuations ($\Phi$) are considered. In the present work, we use the static version of TCM1, including the non-local terms.

The equation that describes the TKE ($\omega$) in TCM1 is given as:
\begin{align}
        \frac{\partial\omega}{\partial t}=\frac{\nabla_{\rm ad} T \Lambda \alpha_{s} c_{p}}{H_{p}^{2}}\sqrt{\omega}(\nabla-\nabla_{\rm ad})-\frac{C_{D}}{\Lambda}\omega^{\frac{3}{2}}-\mathcal{F}_{\omega},
        \label{eqconvflux1}
\end{align}
where $\nabla$ and $\nabla_\mathrm{ad}$ represent the model and adiabatic temperature gradient, respectively. 
$\alpha_{s}$ is a free parameter.
The variable $T$ corresponds to the stellar structure temperature and $c_{p}$ denotes the specific heat capacity at constant pressure.
The viscous dissipation is described by the second term containing the free parameter $C_{D}$.
$\Lambda$ is the length scale of TKE dissipation, which, in the original version of TCM1, is the product of the pressure scale height $H_{p}$ and a free parameter $\alpha_{\Lambda}$, which is analog to the mixing length parameter in MLT. The expression for $\Lambda$ was modified by \citet{wuch95} to account for small convective cores where $H_{p}$ diverges, and reads in the current version of TCM1

\begin{align}
\frac{1}{\Lambda}=\frac{1}{\alpha_{\Lambda}H_{p}} + \frac{1}{\beta r},
\end{align}
where $r$ is the local radius and $\beta$ is a free parameter.

The third term in Eq.~(\ref{eqconvflux1}) describes the non-local fluxes modelled as:
\begin{align}
\mathcal{F}_{\omega} & = -\frac{1}{\overline{\rho}}\nabla\cdot\,\left(\alpha_\omega\,\overline{\rho}\,\Lambda\sqrt{\omega}\,\nabla\,\omega\right).
\end{align}
This equation involves another free parameter $\alpha_{\omega}$ which is used to adjust the measure of non-locality in the model and $\rho$ is the stellar density. The quantities with a bar are the spherically averaged quantities.
As we are interested in the static case, $\frac{\partial \omega}{\partial t}$ is set to zero in Eq.~(\ref{eqconvflux1}).

The main difference between TCM1 and TCM3 is that in the derivation of TCM1 the convective flux is assumed to be proportional to the entropy gradient:
\begin{align}
\Pi=&-\alpha_{s}\Lambda\sqrt{\omega}\frac{\partial s}{\partial r},
\end{align}
where the specific entropy gradient is defined as
\begin{align}
\frac{\partial s}{\partial r}=&-\frac{c_{p}}{H_{p}}(\nabla-\nabla_{\mathrm{ad}}).
\end{align}

Therefore, the temperature gradient from the TCM1 can be obtained using the following equation:
\begin{align}
\nabla - \nabla_{\rm ad} = (\nabla_{\rm rad} - \nabla_{\rm ad}) \left(1 + \frac{\rho c_{p} \alpha_{s} \Lambda \sqrt{\omega}}{\kappa_{\rm rad}}\right)^{-1},
\end{align}
where $\nabla_{\rm rad}$ represents the temperature gradient when all the energy is transported by radiation and $k_{\rm rad}$
is the radiative conduction coefficient.

We refer to the free parameters involved in the TCM1; $\alpha_{\Lambda}$, $\alpha_{\omega}$, ${\alpha_{s}}$, $C_{D}$ and $\beta$ as TC parameters. The $\alpha_{\Lambda}$ is determined by a solar model calibration. A solar calibration is used to find a model with the same radius, luminosity, and surface metal to hydrogen fraction $\frac{Z_{\odot}}{X_{\odot}}$ as the present-day Sun. A sequence of stellar models with varying initial values of $Y$, $Z$ and $\alpha_\mathrm{MLT}$ or $\alpha_{\Lambda}$  is calculated and evolved until the age of the Sun to find the correct combination of these parameters for a solar model. 
The solar calibration was done for a model with MLT with overshooting and a model with the TCM1 using the solar abundances as given by \citet{grev98}. This gave $\alpha_\mathrm{MLT}$=1.79 and $\alpha_{\Lambda}$=1.78 for the model with MLT and the model with the TCM1, respectively. We note that while the $\alpha_\mathrm{MLT}$ is not a part of TCM1, it is needed for the MLT evolutionary 
models which are used for comparison purposes (see Sect.~\ref{sec:ev_models}). $\alpha_{\omega}$ is set to 0.3 following \citet{ahlb22} who found that this value resulted in a similar convective core size as that predicted by a model with MLT including ad hoc overshooting for a $5\,\msun$ star. $\alpha_{s}$ and $C_{D}$ are kept as in \citet{kuhf86,kuhf87} which have been obtained by calibrating the convective flux and convective velocity of the static and local version of TCM1 with the MLT predictions \citep{kuhf87,ahlb22b}. The value $\beta$ is set to 1 following \citet{stra05}. 
The values used in the present work are listed in Table~\ref{table:tab_tc}. For detailed derivations and symbol definitions, we refer the reader to the works of \cite{kuhf86, kuhf87, flas03, kupk22, ahlb22,tere24}.

\begin{table*}
	\centering
	\caption{Values of turbulent convection model parameters used in this study.}
	\label{table:tab_tc}
	\begin{tabular}{lcccccr} 
		\hline
        Parameters & Values & Physical meaning & Source/Reference \\ \hline
        $\alpha_{\Lambda}$ & $1.78$ & Turbulent length scale & Solar calibrated  \\
        $\alpha_{\omega}$ & $0.3$ & Non-locality & \citet{ahlb22}  \\
        $\alpha_{s}$ & $\frac{1}{2}\sqrt{\frac{2}{3}}$ & Entropy flux & \citet{kuhf86,kuhf87}  \\
        $C_{D}$ & $\frac{8}{3}\cdot\sqrt{\frac{2}{3}}$ & Viscous dissipation & \citet{kuhf86,kuhf87}  \\
        $\beta$  & 1.0 & Limitation of $\Lambda$ &  \citet{stra05}  \\
      \hline
	\end{tabular}
\end{table*}

\subsection{Stellar evolutionary models}
\label{sec:ev_models}
In this work, we compute two different sets of models. 
We first compare the evolutionary tracks with different convective approaches in Sect.~\ref{sec:3.1} and then investigate how the characteristics of the evolutionary tracks change if non-locality is included only at specific boundaries (Sect.~\ref{sec:3.2}). 
The second set of models is computed to compare with the observations of 
the detached eclipsing binaries (Sect.~\ref{sec:3.3}). 
For the first set, we computed evolutionary tracks for a star of 
mass $5\,\msun$ from the zero-age main sequence (ZAMS) up to the end of the core He-burning phase.  The tracks were 
computed using MLT alone, using MLT with exponential overshooting as outlined 
in \citet{frey96}, and using the TCM1. 
The initial model for the 
TCM1 is selected from the MLT plus exponential overshooting 
evolutionary tracks at the onset of the main-sequence phase following \citet{ahlb22}.

We adopted the Free Equation of State (FreeEOS) \href{https://freeeos.sourceforge.net/}{(Alan W. Irwin)} and employed OPAL 
opacities \citep{igle96}, supplemented by low-temperature opacities from \citet{ferg05}. For the results presented in Sect.~\ref{sec:3.1} and \ref{sec:3.2}, we have chosen the mass fractions of hydrogen $X=0.7144$, helium $Y=0.2671$, and metals $Z=0.0185$ for the MLT models and $X=0.7152$, $Y=0.2665$, and $Z=0.0183$ for the TCM1 models as obtained from solar calibration.
The overshooting in the models using MLT plus ad-hoc overshooting is controlled by the parameter $f_\mathrm{OV}$, for which we set it to $f_\mathrm{OV}=0.018$. This parameter was calibrated by fitting GARSTEC-isochrones to the color-magnitude diagrams of open cluster \citep{magi10}.
For the MLT and MLT plus overshoot calculations, the calibrated solar 
mixing length parameter is $\alpha_{\rm MLT}= 1.79$, while for the TCM1 approach, it is $\alpha_{\Lambda}=1.78$ (as described in 
Sect.~\ref{sec:2.1}). 

For the results presented in Sect.~\ref{sec:3.3}, we used the same $\alpha_{\Lambda}$. The parameters different from the first set-up are the initial masses and chemical abundances, according to observational results. We will specify them in the corresponding section specific to each system. To convert $\rm{[Fe/H]}$ to $X,Y,Z$, we have used the following relations:
\begin{align}
[\rm{Fe/H}]=&\log{(Z/X)}-\log{(Z/X)}_{\odot},\\
Y= & Y_{0}+\frac{\Delta Y}{\Delta Z}\; Z ~~\mathrm{and} \\ 
X=&1-Y-Z,
\end{align}
where, we have adopted primordial helium $Y_{0}=0.2485$ from \citet{hins13},  ${\frac{Z_{\odot}}{X_{\odot}}=0.0231}$ from \citet{grev98} and He-enrichment ratio $\frac{\Delta Y}{\Delta Z}=0.98$ obtained from solar calibration. We have tabulated all the [Fe/H] and the corresponding $X,Y$ and $Z$ values used in this work in Appendix~\ref{sec:feh}.


\section{Comparison to MLT}
\label{sec:3.1}
\begin{figure}
\centering
\includegraphics[width=0.49\textwidth,keepaspectratio]{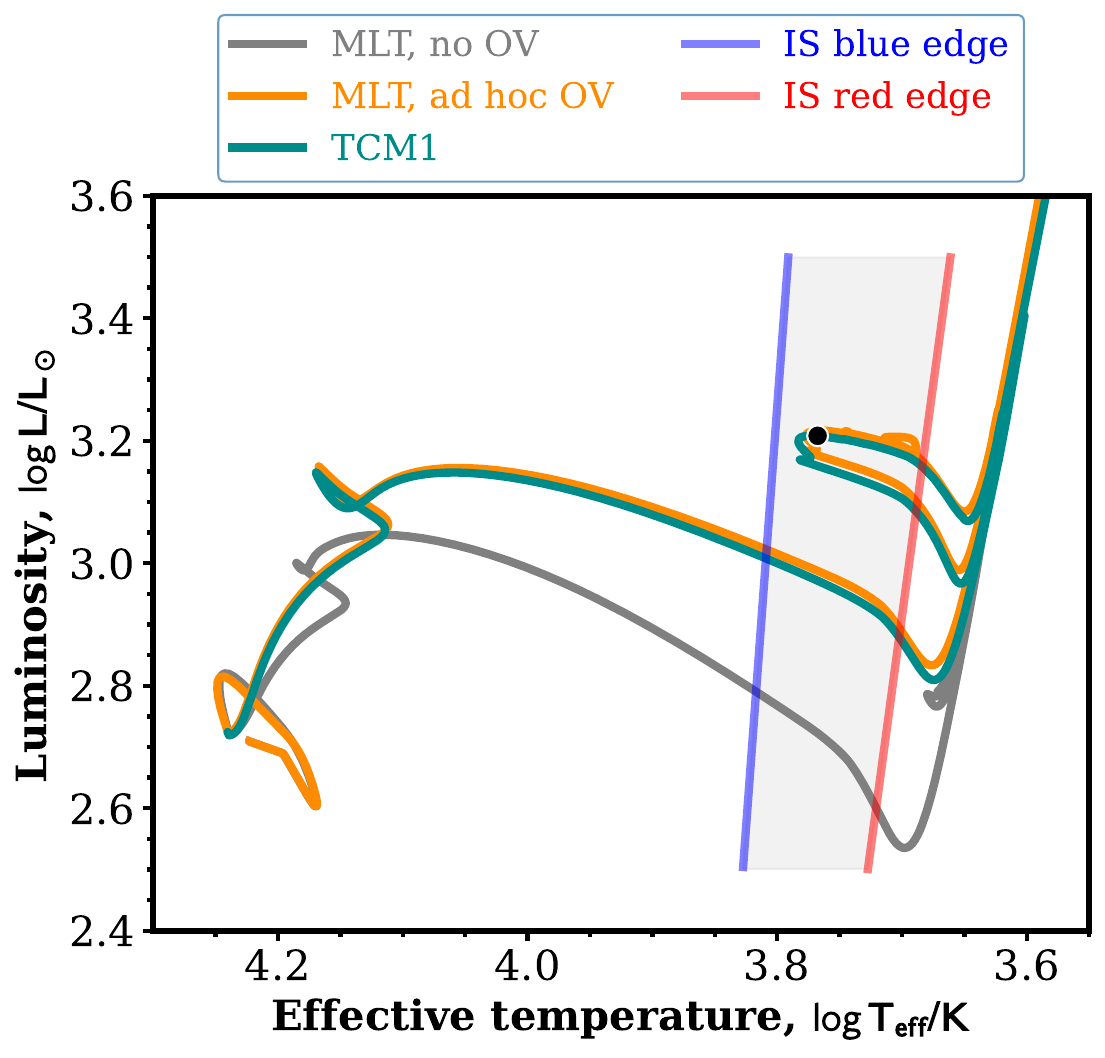}
\caption{Comparison of the evolutionary tracks for a $5\,\msun$ star from the Zero-Age Main Sequence to the core He-burning phase. The tracks are computed using GARSTEC with three different convection treatments: MLT only (grey), MLT with ad-hoc overshooting (orange), and Kuhfuss 1-equation TCM (cyan). The blue and red lines represent the theoretical blue and red edges of the instability strip, respectively, as estimated by \citet{deka24}. The black dot represents a model with a central helium abundance of 0.6, whose turbulent kinetic energy profile is shown in Fig.~\ref{fig:tke}.}
\label{fig:all_hrd_5M}
\end{figure}

The evolutionary tracks for a $5\,\msun$ star calculated with only MLT, with MLT including ad-hoc overshooting, and with
TCM1 are presented in Fig.~\ref{fig:all_hrd_5M}, highlighting the differences among the three tracks. The \ceph\  blue and red instability strip (IS) edges plotted in Fig.~\ref{fig:all_hrd_5M} are taken from \citet{deka24}. It becomes apparent that 
the MLT approach alone fails to reproduce the blue loops of \cephs\ of this (and lower) mass. When including ad-hoc overshooting the blue loops re-appear as can be seen from the orange line in Fig.~\ref{fig:all_hrd_5M} as expected \citep[e.g.][]{moro12}. 
Remarkably, the TCM1 has accomplished this without necessitating fine-tuning of the other involved numerical parameters.
We also note that some other input physics, such as nuclear reaction rates, rotation, and metallicities influence the appearance of the blue loop \citep[e.g.,][]{laut71,robe71,iben72,robe72,bono00,xu04,ande14,zhao23}.
Due to the larger cores on the main sequence, the overall
luminosity increases for the tracks with MLT plus ad-hoc overshooting and 
TCM1 as compared to that with MLT only, thereby resolving the 
mass discrepancy. The extent of the blue loop with TCM1 without fine-tuning the free parameters involved is also similar to that of MLT with ad-hoc overshooting models. Achieving extended blue loops that cross the instability strip for stars with masses closer to $4\,\msun$ and a Galactic composition is very difficult. However, observationally, several such Cepheids have been observed \citep[\eg][]{gall18}. 

\begin{figure}
\centering
\includegraphics[width=0.49\textwidth,keepaspectratio]{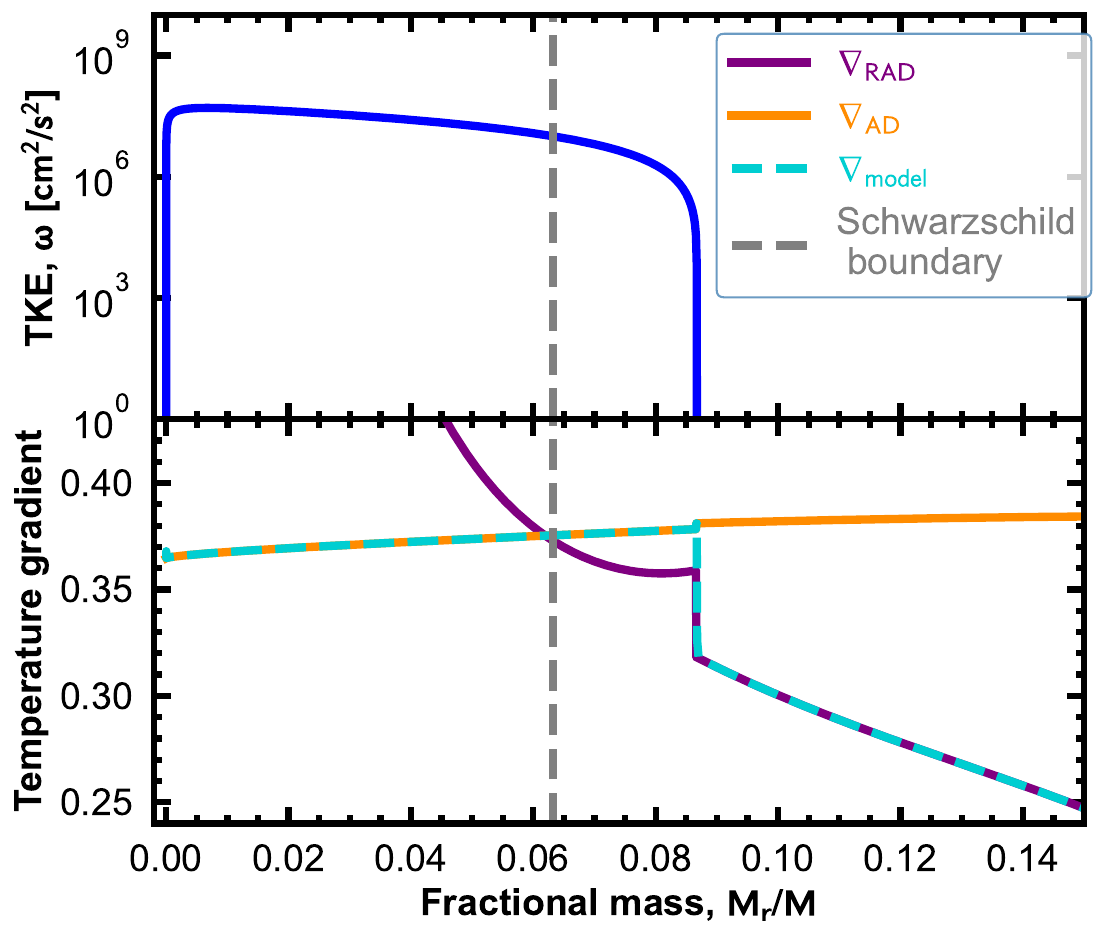}
\caption{The upper panel shows the turbulent kinetic energy profile as
a function of fractional mass  for an evolutionary model of a $5\,\msun$ star computed
using the TCM1 with a central helium abundance of $Y_c=0.6$. 
The lower panel shows the temperature gradient of the model ($\nabla_{\rm model}$, blue dashed line). The purple and orange lines indicate
the radiative $\nabla_{\rm rad}$ and adiabatic $\nabla_{\rm ad}$ temperature gradients, respectively. The vertical grey dashed line denotes the Schwarzschild boundary.}
\label{fig:tke}
\end{figure}

In Fig.~\ref{fig:tke}, we show the internal structure of a $5{\,\msun}$ star with a convective, He-burning core computed with TCM1. This model is marked in Fig.~\ref{fig:all_hrd_5M} by a black marker and has a central helium abundance of $Y_{c}=0.6$. The upper panel of Fig.~\ref{fig:tke} displays the TKE profile as a function of fractional mass. 
The lower panel shows the temperature gradients of the 
model and the
Schwarzschild boundary, where $\nabla_{\rm ad}= \nabla_{\rm 
rad}$. It can be seen that the convective core is extended beyond the 
Schwarzschild boundary, which has been achieved without any additional prescription of overshooting. The temperature gradient does not immediately drop to the radiative gradient at the Schwarzschild boundary. It becomes sub-adiabatic at the Schwarzschild boundary, but it stays very close to the adiabatic value throughout the CBM region and only transitions to the radiative gradient just before reaching the outer boundary of the CBM region.

\section{Influence of CBM at different convective boundaries on \ceph\ blue loops}\label{sec:3.2}
The morphology and extent of the \ceph\ blue loop are influenced by several factors, such as convective core overshooting \citep{cass11}, rotation \citep{ande14, zhao23}, nuclear reaction rates \citep{hala12}, and helium and metal content \citep{bono00}. In this paper, we primarily focus on the role of convection. The significance of convective core overshooting during MS evolution is well established in the literature \citep[e.g.,][]{huan83, cass11}. Convective core overshooting not only increases the MS lifespan but also produces a larger convective core during the core He-burning phase, thereby resulting in a higher luminosity consistent with observations of \cephs\ \citep{huan83}.  

In this section, we discuss the effect of CBM at different boundaries during the core He-burning phase on the blue loops using the Kuhfuss TCM1. The CBM emerges naturally in the solution of the equations due to the non-local nature of the model. To test the effect of non-locality during the core He-burning phase, we first evolved a $5\,\msun$ model until the end of the MS with non-locality employed in the whole star. Then, at the end of the MS, we evolved the model considering non-locality\footnote{for simplicity, we use the term ``non-local'' as a short-cut for ``non-local model for convection in \ldots'' } at all convective boundaries (case I), only at either the boundary of the envelope (case II) or the core (case III) and with no non-locality at all boundaries in the He-burning phase (case IV) as given in Table~\ref{table:CBMcases}.

\begin{table}[t]
\caption{Summary of the cases with CBM at different boundaries during the core He-burning phase}
\centering
\begin{threeparttable}
\label{table:CBMcases}
\begin{tabular}{lcc}
\hline
  & \textbf{core} & \textbf{envelope} \\
\hline
Case I   & non-local & non-local  \\
Case II  & non-local & local      \\
Case III & local     & non-local  \\
Case IV  & local     & local      \\
\hline
\end{tabular}
\end{threeparttable}
\end{table}

\begin{figure*}
\centering
\includegraphics[width=0.9\textwidth,keepaspectratio]{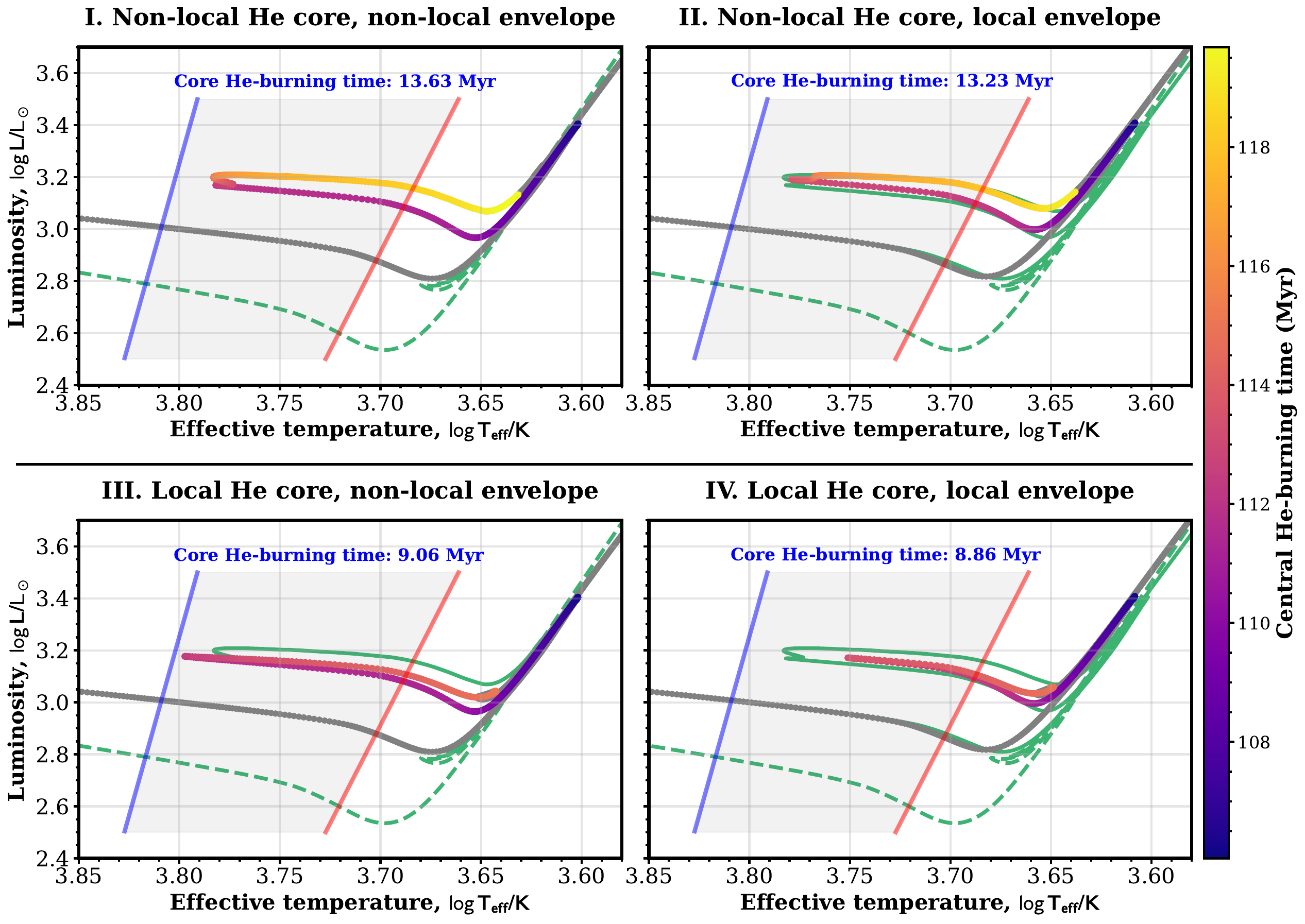}
\caption{Influence of convective boundary mixing at different boundaries during the core He-burning phase on the blue loops of a $5\,\msun$ star. During the previous central hydrogen-burning phase the non-local TCM1 was applied. The four cases I-IV are indicated above the corresponding panel. The blue and red lines represent the blue and red edges of the instability strip, respectively, as estimated by \citet{deka24}. The core He-burning region is color-mapped by age. The green solid and dashed lines indicate the evolutionary tracks with and without non-locality applied to the entire evolution of the star, respectively.}
\label{fig:blueloop_comp}
\end{figure*}

Cases I, II, and III result in similar extensions of the blue loop, while Case IV leads to a shorter blue loop, as shown in 
Fig.~\ref{fig:blueloop_comp}. From Fig.~\ref{fig:blueloop_comp}, it is evident that the timescale of core He-burning is affected by the consideration of the non-locality of convection which also influences the structure of the blue loop. Models with a non-local He core have a longer core helium burning lifetime, as they have a larger reservoir of He to burn through. 
Furthermore, all these different cases seem to slightly affect the positioning of the second and third crossing of the blue 
loop in terms of luminosity.
Blue loops are also generated without non-locality during the core He-burning phase either in the core or in the envelope. This highlights the importance of including non-locality in the MS. However, its inclusion either in the core or in the envelope or the whole star during the core He-burning phase has an impact on the extension and the vertical gap between the second and third crossings of the blue loop. 

The cases shown in Fig.~\ref{fig:blueloop_comp} also allow us to investigate the influence of the non-local treatment of envelope convection (comparing the left and right panels). This treatment influences the extent and efficiency of the first dredge-up, during which the convective envelope penetrates progressively deeper into the former core region, mixing nuclear products of CNO-burning to the surface. In the non-local case, we observe a slightly higher helium enhancement, a more pronounced depletion of $^{12}{\rm C}$, and a lower $^{12}{\rm C}/^{13}{\rm C}$ ratio. However, the differences compared to the local case remain small. For instance, the surface helium abundance increases by 0.025 in the local model and by 0.029 in the non-local one. Likewise, the non-local model exhibits a slightly lower effective temperature, but both evolutionary tracks overlap to a large extent, particularly in terms of the blue loop extension. This relatively small impact is a consequence of the truncated red giant branch (RGB) evolution in intermediate-mass stars. In contrast, for low-mass stars, one observable effect of overshooting at the base of the convective envelope is the shift in the brightness of the red giant bump in stellar clusters \citep{TroisiBump:2011,JoyceBump:2015,FuBump:2018}.

\section{Comparison to observations}
\label{sec:3.3}
For a more comprehensive comparison between the evolutionary model computed using the TCM1 and observations, we selected 
five detached eclipsing binary systems from \citet{pile18}. These systems consist of at least one companion which is a \ceph. Because 
these systems are detached \citep{pile18}, single star evolutionary tracks are in principle suitable for modeling the individual components.
The stellar parameters are obtained from binary modeling, and are listed in Table~\ref{table:tab2} \citep[cf.][]{pile18}.
The input parameters adopted for each system are mentioned in their corresponding subsections. We chose slightly higher initial masses than their current dynamical masses for all the systems. We utilized the same
radiative mass-loss rates from \citet{reim75,reim77}, employing a factor of $\eta=0.2$ for all the systems. The metallicity has a substantial impact on the surface properties of \cephs\ (discussed in Sect.~\ref{sec:discussion}). Unfortunately, precise metallicity measurements are not available for any of the systems discussed in this work. Most previous studies addressing the mass discrepancy problem for OGLE-LMC-CEP-0227 adopted a metallicity of [Fe/H] = -0.33 \citep{cass11,moro12}, based on the mean metallicity of the LMC. However, this value was later refined by \citet{roma22} to $-0.409$, similar to the value by \citet{choud2021} of $-0.42\pm 0.04$.
In fact, \citet{marc13} reported a significantly lower value of [Fe/H] = $-0.65$ for OGLE-LMC-CEP-0227. Therefore, for all systems analyzed in this work, we initially assumed a metallicity of [Fe/H] = -0.5  following \citet{gier15} and \citet{pile18}, changing it to lower or higher values, if we thought this could improve the best fitting model. OGLE-LMC-CEP-1812 was not explored with alternative metallicity values, as its complex evolutionary discrepancies are unlikely to be resolved through adjustments in metallicity alone.

We have estimated the final parameters from the tracks by minimizing the $\chi^2$ statistic, defined as:

\[
\chi^2 = \sum_{i=1}^{N} \frac{\left(O_i - E_i\right)^2}{\sigma_i^2},
\]

where
\begin{itemize}
  \item $O_i$: Observed values of $L / {L_{\odot}}$, $T_{\rm eff}$, and $R / \mathrm{R_{\odot}}$ for each component of the binary systems,
  \item $E_i$: Corresponding theoretical values predicted by the evolutionary tracks,
  \item $\sigma_i$: Uncertainty associated with each observed parameter,
  \item $N$: Total number of observational constraints.
\end{itemize}
It is also important to note that in certain cases, despite achieving a minimum $\chi^2$ value, the derived parameters may seem to deviate from the evolutionary tracks in the HRD due to the small observed radius uncertainty dominating the $\chi^2$. The 
results are summarized in Table~\ref{table:tab_0227}, \ref{table:tab_1812}, \ref{table:tab_4506}, \ref{table:tab_2532}, and \ref{table:tab_1718}.
Note that we do not aim for the absolutely best-fitting model. Our emphasis lies on the comparison between reasonably well-fitting models employing TCM or MLT plus overshooting. As such we did not, for example, vary masses within the observed uncertainty ranges or tested more than two metallicity values.

\subsection{OGLE-LMC-CEP-0227}
\label{sec:ogle227}
OGLE-LMC-CEP-0227 is the first classical \ceph\ that was spectroscopically confirmed to be a member of an eclipsing binary
system, and for which the dynamical masses and other physical parameters were determined \citep{piet10}. 
They found the mass of the \ceph\  (the ``primary'') and the non-pulsating companion (``secondary'') to be very similar ($4.14\pm0.04\,\mathrm{M_\odot}$, resp.\ $4.14\pm0.07\,\mathrm{M_\odot}$), such that within the errors the mass hierarchy is not conclusive. When modelling this system, \citet[][]{cass11} and \citet{moro12} assumed  that the secondary has a slightly higher mass.  
\citet{pile13} estimated the fundamental properties of this system with significantly improved precision as compared to \citet{piet10}. 
In their re-analysis of the orbital solution, they claim that the mass of the companion is indeed lower than that of the \ceph\  ($4.165\pm0.003\,\mathrm{M_\odot}$ vs.\ $4.134\pm0.004\, \mathrm{M_\odot}$). In \citet{pile18} they reported the masses to be $4.15\pm0.03\, \mathrm{M_\odot}$ and $4.06\pm0.03\,\mathrm{M_\odot}$ for the \ceph\ and its companion.
We adopted the values by \citet{pile13} for a first set of models (see Table~\ref{table:tab_0227}).

\begin{figure*}[h!]
\includegraphics[width=1.0\textwidth,keepaspectratio]{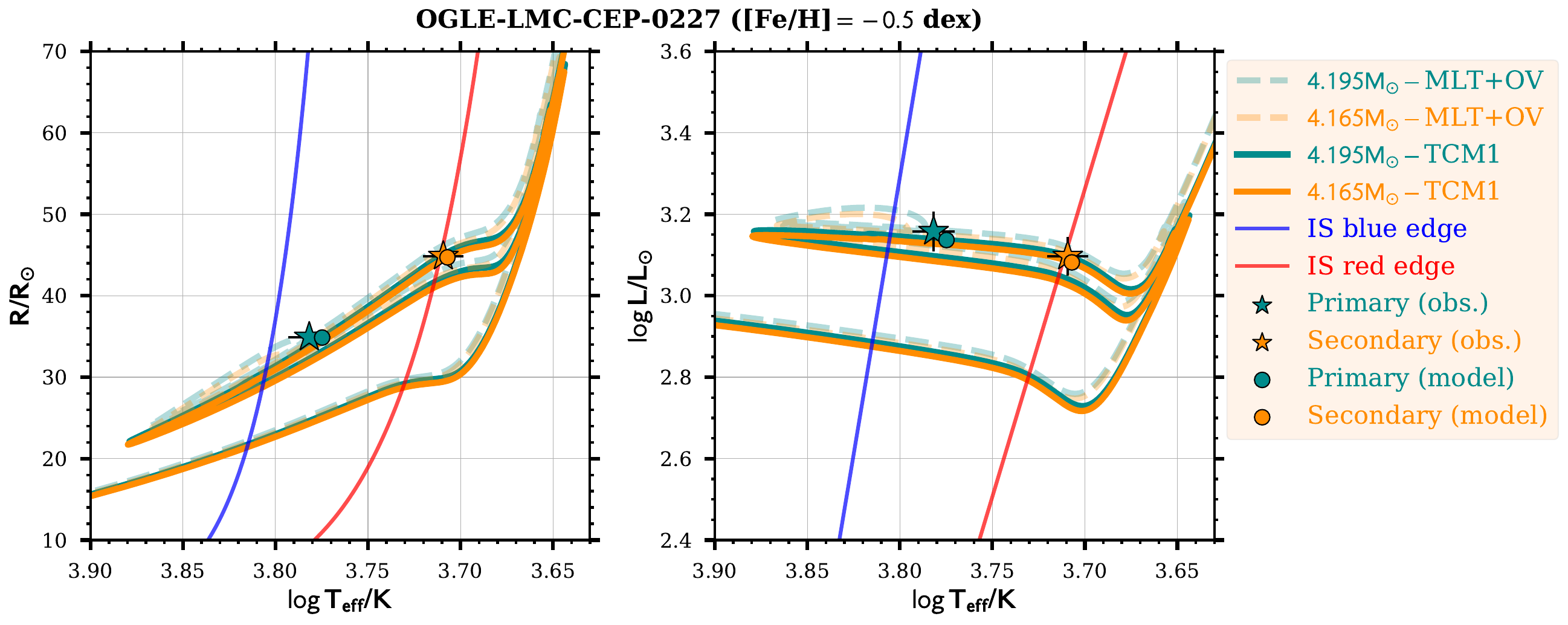}
\caption{The evolutionary tracks for OGLE-LMC-CEP-0227 computed using both MLT plus ad-hoc overshooting (dashed lines) and TCM1 (solid lines). The left panel shows the tracks in the
radius–effective temperature ($RT_{\rm eff}$) plane,
while the right panel shows them in the luminosity–effective temperature ($LT_{\rm eff}$) plane. The observed positions in each plane are indicated by the colored stars (green for the primary, orange for the secondary).
The best-fitting models are shown by colored circles. The observational error bars are indicated in this and all other plots, but are often comparable to or even smaller than the symbols, in particular in the ($RT_{\rm eff}$) plane. The blue and red lines in this plot represent the first overtone \ceph\ blue and red edges taken from \citet{deka24}. This figure is discussed in detail in Sect.~\ref{sec:ogle227}.}
\label{fig:ogle0227_0.5}
\end{figure*}

The adopted initial masses are $M_{p}=4.195\,\msun$ and $M_{s}=4.165\,\msun$ for the primary and secondary, respectively, in order to account for the mass lost in earlier evolutionary phases. We first adopted for these calculations a metallicity of $\rm{[Fe/H]}=-0.5$. The resulting tracks
are shown in Fig.~\ref{fig:ogle0227_0.5}. The left panel shows the tracks in the $R-T_{\rm eff}$, and the right
panel in the $L-T_{\rm eff}$ plane. Both sets of evolutionary tracks consistently predict effective temperatures, 
luminosities, and radii that agree with the observed values. The \ceph\  component occupies the \ceph\  IS, 
whereas the non-pulsating companion is outside the IS. 
 
While our models nicely reproduce the location in the HRD for both components, 
with the selected mass values the evolutionary state of the stars is inconsistent because the secondary (of lower mass, but same age) appears to be further evolved. This contradicts basic stellar physics, and was probably the reason why the authors of previous studies used an inverted mass ratio. 
We followed this approach as well and assigned tentatively the higher mass to the secondary and the lower one to the \ceph. The resulting evolutionary tracks are shown in Fig.~\ref{fig:ogle0227_0.5_alt}. The parameters of the models are also given in Table~\ref{table:tab_0227} under the label of ``model with inverted mass ratio''. Now, the locations along the tracks are consistent with the more massive star being already in a more advanced phase of evolution. We suggest further mass determinations to clarify this relevant issue.

Since this, as all other systems we considered, lacks a
precise metallicity estimate, we further computed evolutionary tracks for this system,   slightly lowering the metallicity to $\rm{[Fe/H]}=-0.6$, keeping the mass assignments as in the first case. 
The parameters of the resulting models that now reproduce the observed locations in the diagrams of Fig.~\ref{fig:ogle0227_0.6} are also given in Table.~\ref{table:tab_0227}, labeled  ``low-metallicity model''. Also in this case, the more massive star is the less evolved component.

An alternative solution to the problem could be of theoretical nature. For the original mass assignments, one could assume that the luminosity of our models might be too low by about 0.1~dex, and therefore both the ad hoc overshooting approach and our TCM1 model are both underestimating the size of the mixed core. In this case (shifting the tracks in Fig.~\ref{fig:ogle0227_0.5} upwards), the \ceph\  would lie on the lower blue loop branch, while the less massive non-pulsating companion would be close to entering it from the red, thereby being in agreement with theoretical expectations for the standard mass hierarchy. 
%
%
This could be achieved by increasing the non-local parameter $\alpha_{\omega}$ of the TCM1 which would increase the convective core size on the MS and in turn lead to a higher luminosity in the core He-burning phase 
\citep[Fig. B.2 in][]{ahlb22}. 

However, 
given the current constraints on the metallicity, it is also difficult to determine whether a different value of $\alpha_{\omega}$ is really needed or whether the mismatch is just an effect of an incorrect metallicity. Considering the assumptions of the TCM1, we expect the same TC parameter values to apply to different systems. 

In conclusion, even this system, which has been modelled several times in the literature, is not free of open questions, still posing challenges to both observers and theorists.

\begin{table}
\caption{The stellar parameters obtained for the components of the OGLE LMC-CEP-0227 system from the evolutionary tracks are listed below. Here, $M$, $R$, $L$, and $T_{\rm eff}$ refer to mass, radius, luminosity, and effective temperature, respectively. The subscripts $p$ and $s$ refer to the primary and secondary components, respectively.}
\centering
\begin{threeparttable}
\label{table:tab_0227}
\begin{tabular}{lcccc}
\hline
\textbf{Parameters} & \textbf{Initial$^{a}$} & \textbf{Final$^{b}$} & \textbf{Observations} \\
\hline
$M_{p}/\msun$ & $4.195$ & $4.166$ & $4.165\pm0.032$ \\
$M_{s}/\msun$ & $4.165$ & $4.136$ & $4.134\pm0.037$ \\
$R_{p}/\rsun$ &  & $34.86$ & $34.92\pm0.29$ \\
$R_{s}/\rsun$ &  & $44.71$ & $44.85\pm0.34$ \\
$\log{(L_{p}/\lsun)}$ &  & $3.136$ & $3.158\pm0.049$ \\
$\log{(L_{s}/\lsun)}$ &  & $3.082$ & $3.097\pm0.047$ \\
$T_{\rm eff,p}$(K) &  & $5953$ & $6050\pm160$ \\
$T_{\rm eff,s}$(K) &  & $5093$ & $5120\pm130$ \\
$\rm age_{p} (Myr)$ &  & $158.18$ &  \\
$\rm age_{s} (Myr)$ &  & $161.13$ &  \\
$\rm{[Fe/H]}$ & $-0.5$ &  &  \\
$\chi_{p}^{2}$ & & 0.76\\
$\chi_{s}^{2}$ & & 0.56 \\
\hline
\multicolumn{4}{c}{``model with inverted mass ratio''}\\
\hline
$M_{p}/M_{\odot}$ & $4.165$ & $4.137$ & $4.165\pm0.032$ \\
$M_{s}/M_{\odot}$ & $4.195$ & $4.166$ & $4.134\pm0.037$ \\
$R_{p}/R_{\odot}$ & & $34.84$ & $34.92\pm0.29$ \\
$R_{s}/R_{\odot}$ & & $44.85$ & $44.85\pm0.34$ \\
$\log{(L_{p}/\lsun)}$ & & $3.124$ & $3.158\pm0.049$ \\
$\log{(L_{s}/\lsun)}$ & & $3.099$ & $3.097\pm0.047$ \\
$T_{\rm eff,p}$(K) & & $5912$ & $6050\pm160$ \\
$T_{\rm eff,s}$(K) & & $5138$ & $5120\pm130$ \\
$\rm age_{p} (Myr)$ & & $160.84$ \\
$\rm age_{s} (Myr)$ & & $158.44$ \\
$\rm{[Fe/H]}$ & $-0.5$ & \\
$\chi_{p}^{2}$ & &1.13 \\
$\chi_{s}^{2}$ & & 0.15 \\
\hline
\multicolumn{4}{c}{``low-metallicity model''}\\
\hline
$M_{p}/\msun$ & $4.195$ & $4.165$ & $4.165\pm0.032$ \\
$M_{s}/\msun$ & $4.165$ & $4.134$ & $4.134\pm0.037$ \\
$R_{p}/\rsun$ &  & $34.96$ & $34.92\pm0.29$ \\
$R_{s}/\rsun$ &  & $44.97$ & $44.85\pm0.34$ \\
$\log{(L_{p}/\lsun)}$ &  & $3.169$ & $3.158\pm0.049$ \\
$\log{(L_{s}/\lsun)}$ &  & $3.129$ & $3.097\pm0.047$ \\
$T_{\rm eff,p}$(K) &  & $6058$ & $6050\pm160$ \\
$T_{\rm eff,s}$(K) &  & $5220$ & $5120\pm130$ \\
$\rm age_{p} (Myr)$ &  & $154.38$ &  \\
$\rm age_{s} (Myr)$ &  & $157.50$ &  \\
$\rm{[Fe/H]}$ & $-0.6$ &  &  \\
$\chi_{p}^{2}$ & & 0.26\\
$\chi_{s}^{2}$ & & 1.09 \\
\hline

\end{tabular}

\begin{tablenotes}
\item[a] Initial parameters considered to compute the evolutionary tracks
\item[b] Obtained parameters from the computed evolutionary tracks
\end{tablenotes}
\end{threeparttable}
\end{table}

\subsection{OGLE-LMC-CEP-1812 }
\label{sec:ogle1812}
\begin{figure*}[htb!]
\includegraphics[width=1.0\textwidth,keepaspectratio]{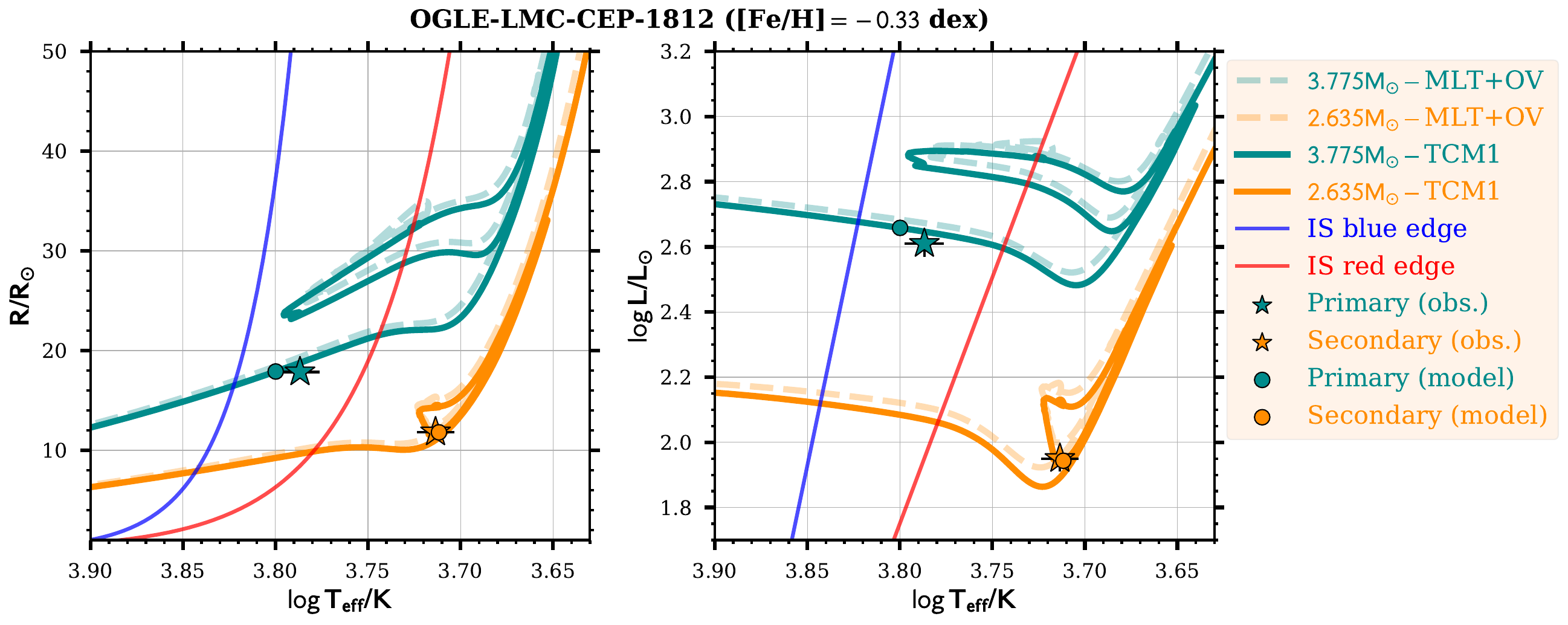}
\caption{As Fig.~\ref{fig:ogle0227_0.5} but for the OGLE-LMC-CEP-1812 system. This figure is discussed in detail in Sect.~\ref{sec:ogle1812}.}
\label{fig:ogle1812}
\end{figure*}

OGLE-LMC-CEP-1812 is the second classical \ceph\  known to exist in an eclipsing binary system. The metallicity is taken to be $\rm{[Fe/H]}=-0.33$ following \citet{piet11}. We adopted the initial masses for the primary and secondary companion 
as $3.775\,\msun$ and $2.635\,\msun$, respectively.  The tracks are displayed in  Fig.~\ref{fig:ogle1812}. Interestingly,
the \ceph\  is found to be in the first crossing of the IS similar to the previous studies \citep[e.g.,][]{neil12}, which is a rare phenomenon  as typically only a few percent of the \cephs\  
should be found at this stage. This is because the intermediate-mass and massive stars evolve on a thermal timescale after they exhaust the hydrogen in their cores \citep[cf.][]{gaut22}. 
Also, another crucial aspect of this system is that, as in the case of OGLE-LMC-CEP-0227, the less massive companion appears to be more evolved than the \ceph, independently of whether it is at the end of the first crossing or already in core He-burning. \citet{neil15} modelled this system using MLT plus
overshooting. They also found that the \ceph\  is located on the first crossing of the IS and that the companion is further evolved than the more massive \ceph \, implying that both stars reach their observed radii at different ages. Indeed, the ages of our best-fitting models differ far more than any uncertainty in the observed parameters could explain. One explanation put forward for this age anomaly is that 
this system originally was a triple system, and the \ceph\  evolved from the merger of two MS stars. Another
possible explanation is that the observed age difference could be attributed to mass transfer between the binary system's two 
components. In both cases, the assumption of single star evolution for both components is unjustified, except when the interaction happened already in the earliest MS phase. More investigation is needed to put constraints on this.
However, we note that the scope of this project is to test the TCM1. We find that again
we obtain the same result with the TCM1 as the typical MLT plus overshooting approach, and that our results are consistent with \citet{neil15}.

\begin{table}[htb!]
\caption{The stellar parameters obtained for the components of the OGLE LMC-CEP-1812 system from the evolutionary tracks are listed below. The symbols have the same meanings as Table~\ref{table:tab_0227}.}
\centering
\begin{threeparttable}
\label{table:tab_1812}
\begin{tabular}{lcccc}
\hline
\textbf{Parameters} & \textbf{Initial$^{a}$} & \textbf{Final$^{b}$} & \textbf{Observations} \\
\hline
$M_{p}/\msun$ & $3.78$ & $3.76$ & $3.76\pm0.03$ \\
$M_{s}/\msun$ & $2.64$ & $2.62$ & $2.62\pm0.02$ \\
$R_{p}/\rsun$ &  & $17.90$ & $17.85\pm0.13$ \\
$R_{s}/\rsun$ &  & $11.80$ & $11.83\pm0.08$ \\
$\log{(L_{p}/\lsun)}$ &  & $2.66$ & $2.61\pm0.04$ \\
$\log{(L_{s}/\lsun)}$ &  & $1.94$ & $1.95\pm0.04$ \\
$T_{\rm eff,p}$(K) &  & $6307$ & $6120\pm150$ \\
$T_{\rm eff,s}$(K) &  & $5147$ & $5170\pm120$ \\
$\rm age_{p} (Myr)$ &  & $186.57$ &  \\
$\rm age_{s} (Myr)$ &  & $471.47$ &  \\
$\rm{[Fe/H]}$ & $-0.33$ &  &  \\
$\chi_{p}^{2}$ & & 1.78\\
$\chi_{s}^{2}$ & & 0.42 \\
\hline

\end{tabular}

\begin{tablenotes}
\item[a] Initial parameters considered to compute the evolutionary tracks
\item[b] Obtained parameters from the computed evolutionary tracks
\end{tablenotes}
\end{threeparttable}
\end{table}

\subsection{OGLE-LMC-CEP-4506 }
\label{sec:ogle4506}
\begin{figure*}[htb!]
\includegraphics[width=1.0\textwidth,keepaspectratio]{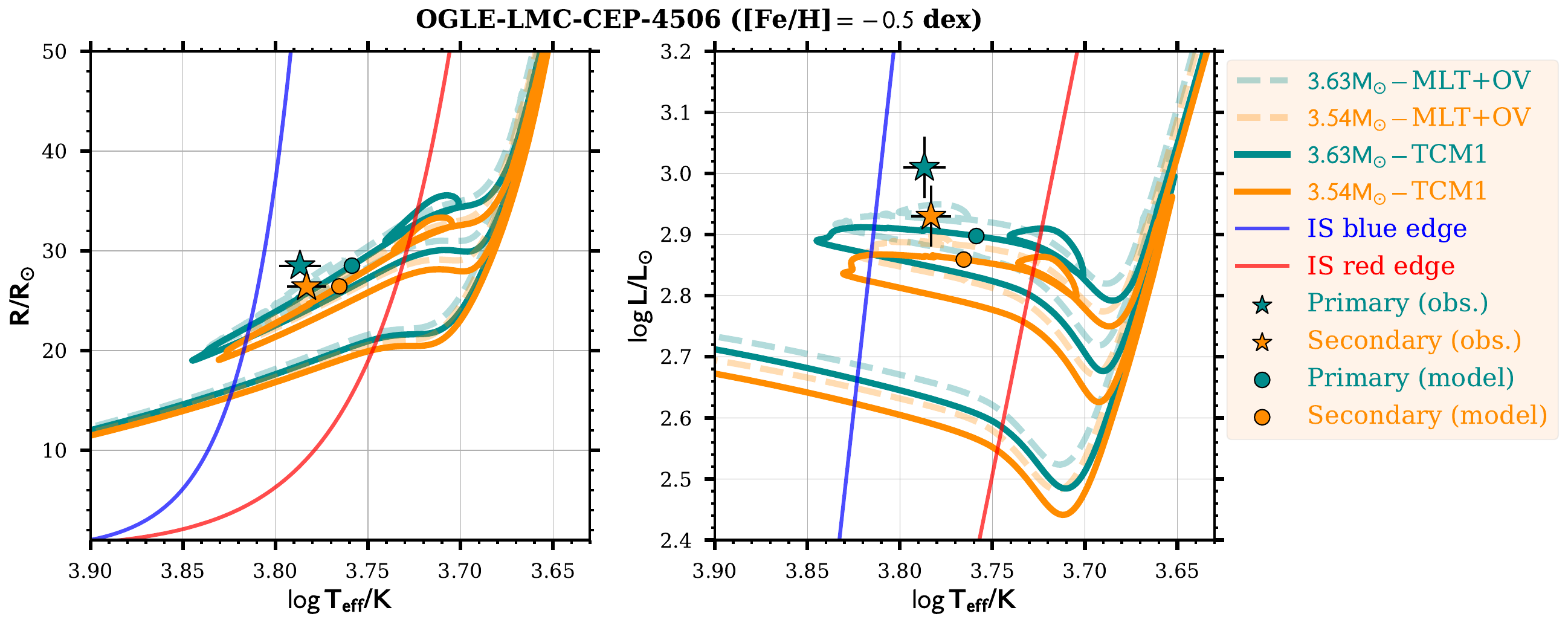}
\caption{As Fig.~\ref{fig:ogle0227_0.5} but for OGLE-LMC-CEP-4506 system with $\rm[Fe/H]=-0.5$. This figure is discussed in detail in Sect.~\ref{sec:ogle4506}.}
\label{fig:ogle4506_0.5}
\end{figure*}

This highly excentric binary system was analysed by \citet{gier15}\footnote{Originally called OGLE~LMC~562.05.9009}. While both companions appear to be in the IS, the secondary with mass  $3.52\,\msun$ is not pulsating. The primary is a slightly more massive \ceph\ of $3.61\,\msun$. Both components have very similar effective temperatures\footnote{The observed effective temperature indicates the mean effective temperature over a pulsation cycle.}, determined by \citet{pile18} to be $6120\pm160$~K and $6070\pm150$~K, for the primary and secondary, respectively.
We have computed the evolution for each of the stars in the OGLE-LMC-CEP-4506 system with initial masses of $3.63\,\msun$ (\ceph) and $3.54\,\msun$ (non-pulsating companion), and  a metallicity [Fe/H] of $-0.5$ (Fig.~\ref{fig:ogle4506_0.5}). 
The slightly more massive primary appears to be located on the upper branch of the blue loop (see in particular the left panel of Fig.~\ref{fig:ogle4506_0.5}), and is therefore more evolved, as to be expected.
\citet{gier15} modelled this system using MLT plus overshooting and found a similar result, but speculate that a better determination of $T_\mathrm{eff}$ could place the secondary outside the red edge of the IS. This appears to be unlikely, as the observed temperature is too high and the later re-determination by \citet{pile18} basically confirmed the earlier values, although, due to the very long orbital period of 1550 days, acquiring well-distributed spectroscopic data presents a significant challenge. Further observations are probably needed for this system to put precise constraints on its stellar parameters  \citep{pile18}. Our chosen metallicity is identical to the one used by \citet{pile18}, although it appears to be too low compared to the average LMC value of $\mathrm{[Fe/H]}=-0.42$ by \citet{choud2021}. The best-fit models from this set of tracks exhibit luminosities that are approximately $\sim 2\sigma$ lower than the observed values. Therefore, we further computed another set of evolutionary tracks with the same initial masses but with an even lower metallicity value of $-0.6$ (Fig.~\ref{fig:ogle4506_0.6}). We obtain a marginally lower $\chi^2$ for this set of calculation, but otherwise the same result. In both cases, the best-fitting models differ by about 6\% in age, which is slightly more than the range allowed by the uncertainty in mass (1\% for both stars).

We note that computations with MLT plus ad-hoc overshooting lead to a slightly better agreement between the evolutionary tracks and the observations (see Fig.~\ref{fig:ogle4506_0.6})  due to a larger convective core size on the MS (see \citealt{ahlb22}, their Fig. 7). This also could be achieved in the TCM model by increasing $\alpha_{\omega}$, as discussed in Sect.~\ref{sec:ogle227}. The difference between the tracks is however much smaller than compared to MLT without overshooting, as mentioned before.

\begin{table}[htb!]
\caption{The stellar parameters obtained for the components of the OGLE LMC-CEP-4506 system from the evolutionary tracks are listed below. The symbols have the same meanings as Table~\ref{table:tab_0227}.}
\centering
\begin{threeparttable}
\label{table:tab_4506}
\begin{tabular}{lccc}
\hline
\textbf{Parameters} & \textbf{Initial$^{a}$} & \textbf{Final$^{b}$} & \textbf{Observations} \\
\hline
$M_{p}/\msun$ & $3.63$ & $3.61$ & $3.61\pm0.03$ \\
$M_{s}/\msun$ & $3.54$ & $3.52$ & $3.52\pm0.03$ \\
$R_{p}/\rsun$ &  & $28.4$ & $28.5\pm0.2$ \\
$R_{s}/\rsun$ &  & $26.2$ & $26.4\pm0.2$ \\
$\log{(L_{p}/\lsun)}$ &  & $2.89$ & $3.01\pm0.05$ \\
$\log{(L_{s}/\lsun)}$ &  & $2.86$ & $2.93\pm0.05$ \\
$T_{\rm eff,p}$(K) &  & $5743$ & $6120\pm160$ \\
$T_{\rm eff,s}$(K) &  & $5850$ & $6070\pm150$ \\
$\rm age_{p} (Myr)$ &  & $222.59$ &  \\
$\rm age_{s} (Myr)$ &  & $236.38$ &  \\
$\rm{[Fe/H]}$ & $-0.5$ &  &  \\
$\chi_{p}^{2}$ & & 3.25\\
$\chi_{s}^{2}$ & & 2.14 \\
\hline
\multicolumn{4}{c}{``low-metallicity model''}\\
\hline
$M_{p}/\msun$ & $3.63$ & $3.61$ & $3.61\pm0.03$ \\
$M_{s}/\msun$ & $3.54$ & $3.52$ & $3.52\pm0.03$ \\
$R_{p}/\rsun$ &  & $28.4$ & $28.5\pm0.2$ \\
$R_{s}/\rsun$ &  & $26.4$ & $26.4\pm0.2$ \\
$\log{(L_{p}/\lsun)}$ &  & $2.94$ & $3.01\pm0.05$ \\
$\log{(L_{s}/\lsun)}$ &  & $2.90$ & $2.93\pm0.05$ \\
$T_{\rm eff,p}$(K) &  & $5891$ & $6120\pm160$ \\
$T_{\rm eff,s}$(K) &  & $5982$ & $6070\pm150$ \\
$\rm age_{p} (Myr)$ &  & $216.98$ &  \\
$\rm age_{s} (Myr)$ &  & $230.23$ &  \\
$\rm{[Fe/H]}$ & $-0.6$ &  &  \\
$\chi_{p}^{2}$ & & 2.04\\
$\chi_{s}^{2}$ & & 0.81 \\
\hline

\end{tabular}

\begin{tablenotes}
\item[a] Initial parameters considered to compute the evolutionary tracks
\item[b] Obtained parameters from the computed evolutionary tracks
\end{tablenotes}
\end{threeparttable}
\end{table}

\subsection{OGLE-LMC-CEP-2532 }
\label{sec:ogle2532}
\begin{figure*}[htb!]\includegraphics[width=1.0\textwidth,keepaspectratio]{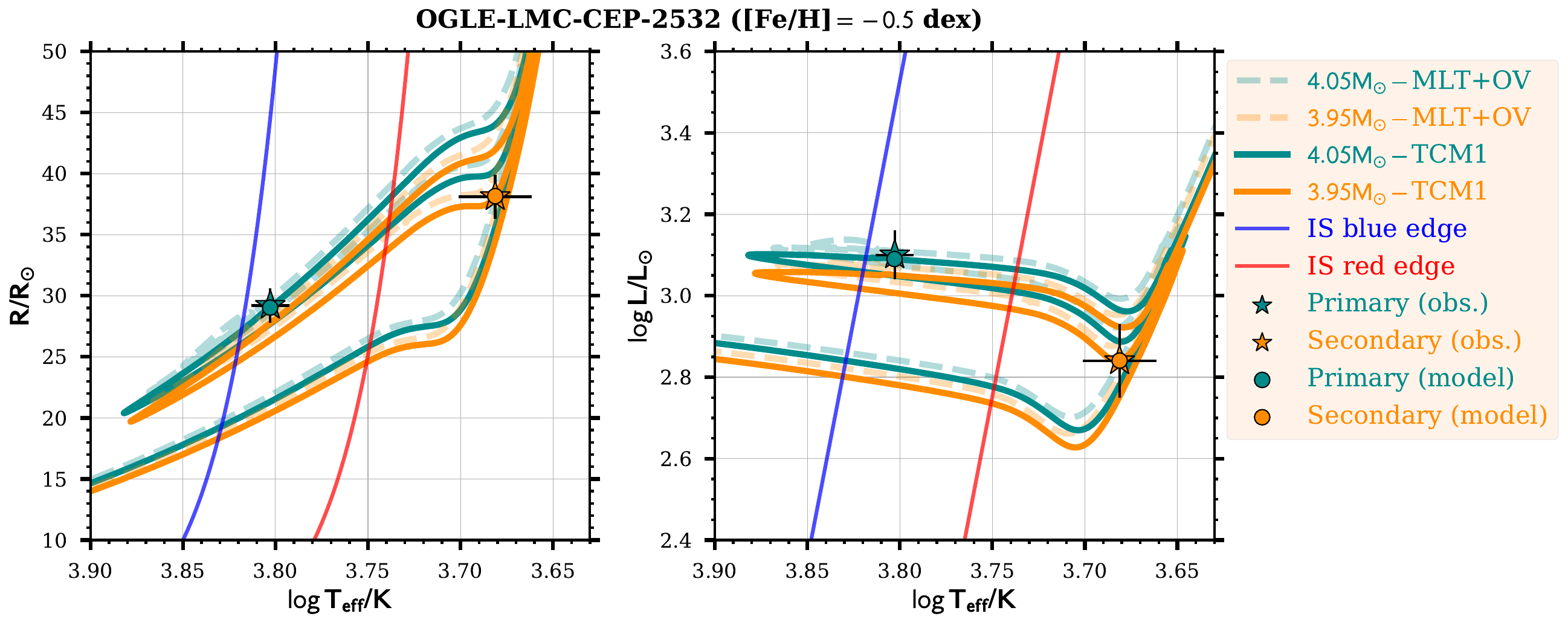}
\caption{As Fig.~\ref{fig:ogle0227_0.5} but for the OGLE-LMC-CEP-2532 system, for a metallicity of $\rm{[Fe/H]=-0.5}$. This figure is discussed in detail in Sect.~\ref{sec:ogle2532}.}
\label{fig:ogle2532_0.5}
\end{figure*}

OGLE-LMC-CEP-2532 is a binary system with a first overtone \ceph\ as the primary component and a non-pulsating star as the secondary component. Its orbit was first analysed by \citet{pile15},  and later revised by \citet{pile18}. We used the latter results for the stellar parameters. The \ceph\ in this system has the most
accurate dynamical mass ($3.98\pm 0.10\,\mathrm{M_\odot}$) determined for a first overtone \ceph. The secondary, non-pulsating companion is slightly less massive ($3.94\pm0.09\,\mathrm{M_\odot}$), although within the errors their masses could also be equal.
We adopted initial masses of $4.05\,\msun$ and $3.95\,\msun$, respectively, and as before two metallicity values: [Fe/H] of $-0.5$ and $-0.6$ (see Tab.~\ref{table:tab_2532}). The tracks corresponding to $-0.5$ are shown in 
Fig.~\ref{fig:ogle2532_0.5}, and those for $-0.6$ are shown in Fig.~\ref{fig:ogle2532_0.6}.  Both stars can be modelled almost exactly in the corresponding $T_\mathrm{eff}$ -- $R/\mathrm{R_\odot}$ and HR diagrams. The primary is already on the third crossing of the blue loop for both sets of models. For the lower metallicity, $\mathrm{[Fe/H]}=-0.6$, the secondary is still on the lower RGB, at the onset of core He-burning. For $\mathrm{[Fe/H]}=-0.5$ it may be in this phase, too, or has already entered the lower branch of the loop. The evolutionary age differences of the two components of the best-fitting models are $\sim 1$  and $7$~Myrs, respectively, for the higher and lower metallicity. This is well within the mass uncertainty. 
The $\chi^2$ value is found to be minimal for  [Fe/H]$=-0.5$.   This is one of the best theory-consistent systems in this work.

\begin{table}[htb!]
\caption{The stellar parameters obtained for the components of the OGLE LMC-CEP-2532 system from the evolutionary tracks are listed below.  The symbols have the same meanings as Table~\ref{table:tab_0227}.}
\centering
\begin{threeparttable}
\label{table:tab_2532}
\begin{tabular}{lccc}
\hline
\textbf{Parameters} & \textbf{Initial$^{a}$} & \textbf{Final$^{b}$} & \textbf{Observations} \\
\hline
$M_{p}/\msun$ & $4.05$ & $4.02$ & $3.98\pm0.10$ \\
$M_{s}/\msun$ & $3.95$ & $3.93$ & $3.94\pm0.09$ \\
$R_{p}/\rsun$ &  & $29.0$ & $29.2\pm1.4$ \\
$R_{s}/\rsun$ &  & $38.1$ & $38.1\pm1.8$ \\
$\log{(L_{p}/\lsun)}$ &  & $3.09$ & $3.10\pm0.06$ \\
$\log{(L_{s}/\lsun)}$ &  & $2.84$ & $2.84\pm0.09$ \\
$T_{\rm eff,p}$(K) &  & $6351$ & $6350\pm150$ \\
$T_{\rm eff,s}$(K) &  & $4799$ & $4800\pm220$ \\
$\rm age_{p} (Myr)$ &  & $171.36$ &  \\
$\rm age_{s} (Myr)$ &  & $171.12$ &  \\
$\rm{[Fe/H]}$ & $-0.5$ &  &  \\
$\chi_{p}^{2}$ & & 0.20\\
$\chi_{s}^{2}$ & & 0.02 \\
\hline
\multicolumn{4}{c}{``low-metallicity model''}\\
\hline
$M_{p}/\msun$ & $4.05$ & $4.02$ & $3.98\pm0.10$ \\
$M_{s}/\msun$ & $3.95$ & $3.94$ & $3.94\pm0.09$ \\
$R_{p}/\rsun$ &  & $29.1$ & $29.2\pm1.4$ \\
$R_{s}/\rsun$ &  & $38.1$ & $38.1\pm1.8$ \\
$\log{(L_{p}/\lsun)}$ &  & $3.09$ & $3.10\pm0.06$ \\
$\log{(L_{s}/\lsun)}$ &  & $2.83$ & $2.84\pm0.09$ \\
$T_{\rm eff,p}$(K) &  & $6349$ & $6350\pm150$ \\
$T_{\rm eff,s}$(K) &  & $4776$ & $4800\pm220$ \\
$\rm age_{p} (Myr)$ &  & $163.86$ &  \\
$\rm age_{s} (Myr)$ &  & $157.38$ &  \\
$\rm{[Fe/H]}$ & $-0.6$ &  &  \\
$\chi_{p}^{2}$ & & 0.14\\
$\chi_{s}^{2}$ & & 0.14 \\
\hline

\end{tabular}

\begin{tablenotes}
\item[a] Initial parameters considered to compute the evolutionary tracks
\item[b] Obtained parameters from the computed evolutionary tracks
\end{tablenotes}
\end{threeparttable}
\end{table}

\subsection{OGLE-LMC-CEP-1718 }
\label{sec:ogle1718}

OGLE-LMC-CEP-1718 consists of two first-overtone (FO) \cephs. It was first detected by \citet{soszy2008} and spectroscopically confirmed by \citet{gier2014}.
It is the first ever detected binary system which consists of two FO \cephs, and presents a unique opportunity to test both evolutionary and pulsation models. Here, the primary component is slightly more massive than the secondary 4.27 vs.\ 4.22~$\mathrm{M_\odot}$, yet both have similar effective temperatures, being within 40~K at an uncertainty of 150~K (see Table~\ref{table:tab_1718}). Interestingly, the secondary component exhibits higher luminosity than the primary, which suggests that the secondary, despite its lower mass, is in a more advanced evolutionary stage. This again contradicts basic stellar physics, where the more massive star is expected to evolve more quickly. However, within the quoted mass uncertainty of $0.04\,\mathrm{M_\odot}$ the mass hierarchy could also be inverted, reconciling observations with theoretical expectations for the evolution of two non-interacting single stars. 

\begin{figure*}[htb!]
\includegraphics[width=1.0\textwidth,keepaspectratio]{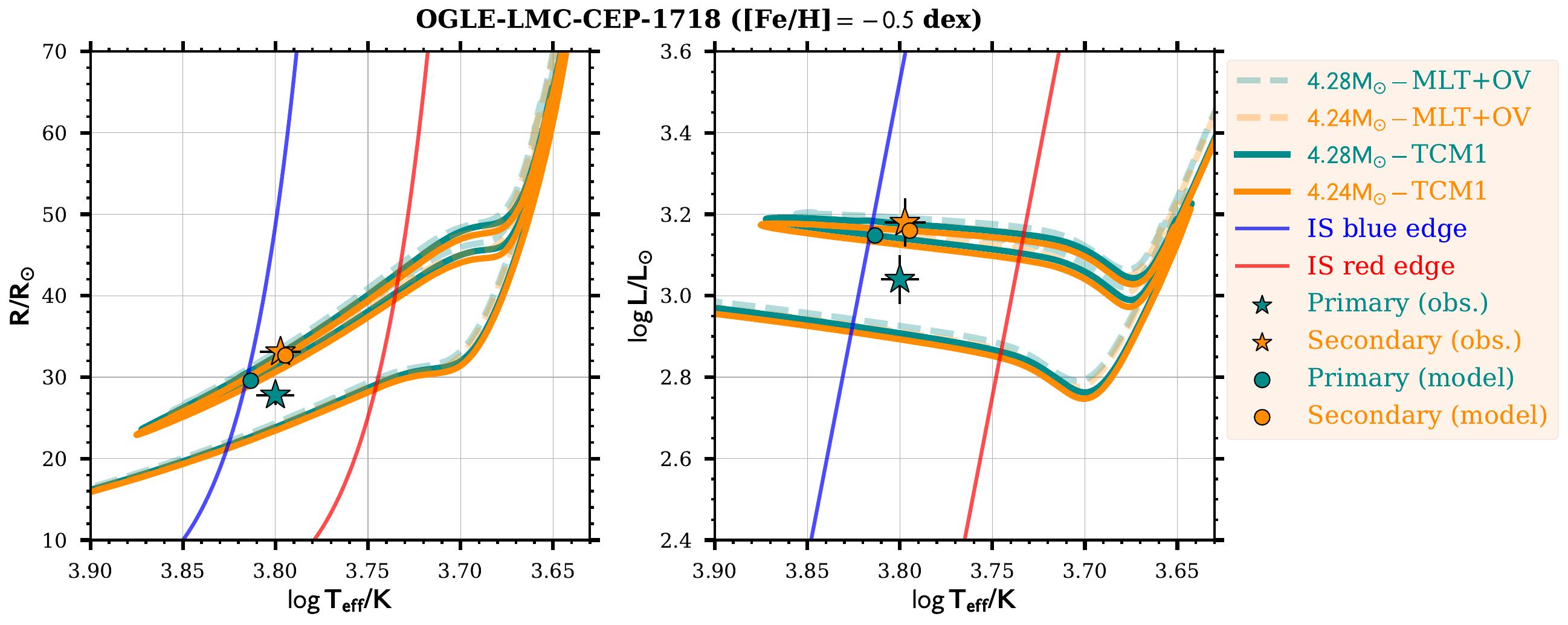}
\caption{As Fig.~\ref{fig:ogle2532_0.5} but for the OGLE-LMC-CEP-1718 system with $\rm{[Fe/H]=-0.5}$. This figure is discussed in detail in Sect.~\ref{sec:ogle1718}.}
\label{fig:ogle1718_0.5}
\end{figure*}

For this system, we have adopted initial masses of $4.28\, \msun$ and $4.24\,\msun$, and considered three metallicity values: [Fe/H] = $-0.5$,$-0.6$ and $-0.3$ (the motivation for the highest value is given below). The evolutionary tracks corresponding to these metallicities are shown in 
Fig.~\ref{fig:ogle1718_0.5},~\ref{fig:ogle1718_0.6} and ~\ref{fig:ogle1718_0.3}, respectively. While the secondary component's observed parameters align well with theoretical predictions with metallicity [Fe/H] = $-0.6$, there is a mismatch in radius and luminosity between the observed and predicted parameters for the primary \ceph. Increasing the metallicity will shift the evolutionary track to lower luminosity thereby reducing the difference between observation and models. The result of this calculation is shown in Appendix \ref{fig:ogle1718_0.3}.  We note that in all three metallicity cases, the observed parameters are recovered within $2\sigma$. However, the age differences are incompatible with the uncertainties in mass for all metallicities, but lowest in the [Fe/H] = $-0.3$ case (Table~\ref{table:tab_1718}).

The radius or luminosity difference between the two objects is too large to assume that they are both in the blue loop phase, but in different crossings, as the lower and upper branches are too close to each other at the objects' temperature. Rather, radii or luminosities suggest that one star is on the first crossing, and the other one in the loop. Furthermore, the secondary is moving blueward, based on period change measurements \citep{pile18}, which is a further indication that the secondary is actually on the second crossing (lower loop branch). Together with a mass inversion, this would yield consistency with theory. For the lowest metallicity, [Fe/H] = $-0.6$, we indeed find that one star is on the first crossing, while the other is in the loop (Fig.~\ref{fig:ogle1718_0.6}). \citet{pile18} also commented on this problem, suggesting physical effects like rotation or overshooting as possible solutions. However, given the very similar masses, such physical effects should be similar in both stars.

These problems may stem from the complex nature of the system, as both components are pulsating variables, making light curve modeling particularly challenging. Additionally, the system presents further difficulty since only one shallow eclipse is visible in each cycle. As a result, the parameters derived from light curve modeling rely on several assumptions and are subject to significant uncertainty.

\begin{table}[htb!]
\caption{The stellar parameters obtained for the components of the OGLE LMC-CEP-1718 system from the evolutionary tracks are listed below.  The symbols have the same meanings as Table~\ref{table:tab_0227}.}
\centering
\begin{threeparttable}
\label{table:tab_1718}
\begin{tabular}{lccc}
\hline
\textbf{Parameters} & \textbf{Initial$^{a}$} & \textbf{Final$^{b}$} & \textbf{Observations}  \\
\hline
$M_{p}/\msun$ & $4.28$ & $4.25$ & $4.27\pm0.04$  \\
$M_{s}/\msun$ & $4.24$ & $4.21$ & $4.22\pm0.04$ \\
$R_{p}/\rsun$ &  & $29.6$ & $27.8\pm1.2$ \\
$R_{s}/\rsun$ &  & $32.6$ & $33.1\pm1.3$ \\
$\log{(L_{p}/\lsun)}$ &  & $3.16$ & $3.04\pm0.06$ \\
$\log{(L_{s}/\lsun)}$ &  & $3.15$ & $3.18\pm0.06$  \\
$T_{\rm eff,p}$(K) &  & $6506$ & $6310\pm150$ \\
$T_{\rm eff,s}$(K) &  & $6231$ & $6270\pm160$ \\
$\rm age_{p} (Myr)$ &  & $147.06$ &  \\
$\rm age_{s} (Myr)$ &  & $154.14$ &  \\
$\rm{[Fe/H]}$ & $-0.5$ &  &  \\
$\chi_{p}^{2}$ & & 2.66\\
$\chi_{s}^{2}$ & & 0.52 \\
\hline
\multicolumn{4}{c}{``low-metallicity model''} \\
\hline
$M_{p}/\msun$ & $4.28$ & $4.27$ & $4.27\pm0.04$ \\
$M_{s}/\msun$ & $4.24$ & $4.21$ & $4.22\pm0.04$ \\
$R_{p}/\rsun$ &  & $25.8$ & $27.8\pm1.2$ \\
$R_{s}/\rsun$ &  & $33.2$ & $33.1\pm1.3$ \\
$\log{(L_{p}/\lsun)}$ &  & $2.92$ & $3.04\pm0.06$ \\
$\log{(L_{s}/\lsun)}$ &  & $3.18$ & $3.18\pm0.06$ \\
$T_{\rm eff,p}$(K) &  & $6127$ & $6310\pm150$ \\
$T_{\rm eff,s}$(K) &  & $6276$ & $6270\pm160$ \\
$\rm age_{p} (Myr)$ &  & $131.24$ &  \\
$\rm age_{s} (Myr)$ &  & $150.06$ &  \\
$\rm{[Fe/H]}$ & $-0.6$ &  &  \\
$\chi_{p}^{2}$ & & 2.75\\
$\chi_{s}^{2}$ & & 0.09 \\
\hline
\multicolumn{4}{c}{``high-metallicity model''}\\
\hline
$M_{p}/\msun$ & $4.28$ & $4.26$ & $4.27\pm0.04$   \\
$M_{s}/\msun$ & $4.24$ & $4.21$ & $4.22\pm0.04$  \\
$R_{p}/\rsun$ &  & $28.9$ & $27.8\pm1.2$ \\
$R_{s}/\rsun$ &  & $32.4$ & $33.1\pm1.3$ \\
$\log{(L_{p}/\lsun)}$ &  & $3.11$ & $3.04\pm0.06$ \\
$\log{(L_{s}/\lsun)}$ &  & $3.13$ & $3.18\pm0.06$ \\
$T_{\rm eff,p}$(K) &  & $6444$ & $6310\pm150$ \\
$T_{\rm eff,s}$(K) &  & $6158$ & $6270\pm160$ \\
$\rm age_{p} (Myr)$ &  & $147.91$ &  \\
$\rm age_{s} (Myr)$ &  & $156.14$ &  \\
$\rm{[Fe/H]}$ & $-0.3$ &  &  \\
$\chi_{p}^{2}$ & & 1.77\\
$\chi_{s}^{2}$ & & 1.18 \\
\hline

\end{tabular}

\begin{tablenotes}
\item[a] Initial parameters considered to compute the evolutionary tracks
\item[b] Obtained parameters from the computed evolutionary tracks
\end{tablenotes}
\end{threeparttable}
\end{table}

\section{Summary and Discussion}
\label{sec:discussion}
\begin{figure*}
	\centering
	\includegraphics[width=1.0\textwidth,keepaspectratio]{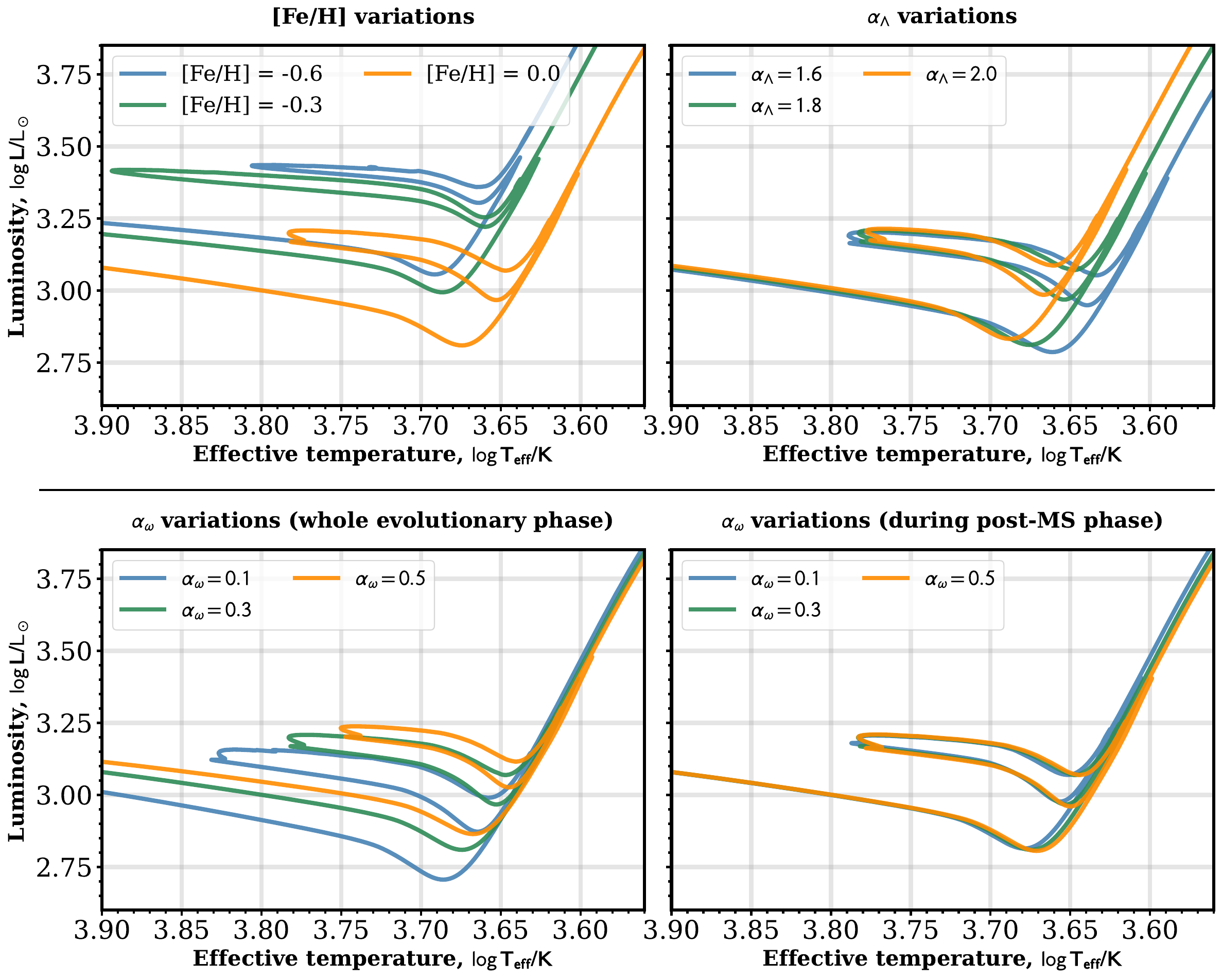}
	\caption{Evolutionary tracks of a $5\;\msun$ \ceph{} for varying metallicity, dissipation length parameter $\alpha_{\Lambda}$ and non-local parameter $\alpha_{\omega}$. The variations of the metallicity and $\alpha_\Lambda$ are shown in the top left and right panel, respectively. In the lower left panel $\alpha_\omega$ is varied in the whole evolution of the star, while in the lower right panel $\alpha_\omega$ is only varied in the post main-sequence evolution. Parameters other than the ones indicated are kept at their default values.}
	\label{fig:paramdep}
\end{figure*}

The evolutionary tracks, particularly the blue loops and luminosity levels of \cephs\ are highly sensitive to variations of the model parameters and assumptions about the input physics. 
As is known and as we have shown, CBM is a crucial process that needs to be included in evolutionary models to successfully model Cepheids, in particular to reproduce the correct luminosity at a given stellar mass. In general, MLT plus overshooting models are adequate in predicting the observations of \cephs. However, the exact amount of overshooting required to explain individual cases remains uncertain and the free parameter in the overshooting prescription needs to be determined.
Our aim in this study was to improve the modeling of convection and to provide a more physically grounded approach to treating non-locality. We have included slight mass loss in our models, but have neither concentrated on varying its amount, nor considered other, potentially influential effects such as rotation.

While mass and radius of the five \ceph\ systems we modelled are known with high precision, their metallicity is rather uncertain. The influence of metallicity is demonstrated in
the top left panel of Fig.~\ref{fig:paramdep}, where we present the evolutionary tracks for a $5\;\msun$ star corresponding to three different metallicity values ($[{\rm Fe/H}] = -0.6, -0.3, 0.0$). Metallicity has a significant impact on the luminosity of the tracks, the temperature along the RGB and the extent of the blue loops, particularly  when we transition from the metal-poor to the metal-rich models (from $[{\rm Fe/H}] = -0.6$ to $0.0$ ). 
These effects mainly originate from the evolution on the main sequence, where more metal-rich stars are cooler and fainter than their metal-poor counterparts, due to the higher opacities. This well-known and global effect pertains throughout the evolution and explains the trends seen in the upper left panel of Fig.~\ref{fig:paramdep}: the RGB parts of more metal-poor models are slightly hotter, loops more extended, and radii somewhat larger. However, the influence of metallicity on the details and extent of the blue loop is not as clear, and not necessarily monotonic \citep[cf.\ their Figs.~6.10 and 6.11]{SalarisCassisi:2005}.
As an example, we refer to Sect.~\ref{sec:ogle1718}, where a higher metallicity than usually assumed leads to a fainter track, locates the \ceph\ component on the second crossing and thereby allows for a better fit  (compare  Figs.~\ref{fig:ogle1718_0.5} and \ref{fig:ogle1718_0.3}).
Understanding the details of the influence of metallicity will require a detailed investigation, which is beyond the scope of this work. Nonetheless, the evolution of \cephs\ during core helium burning is obviously highly sensitive to their metallicity.

We also tested the influence of the parameters of TCM1 (see Sect.~\ref{sec:2.1}).
In the top right panel of Fig.~\ref{fig:paramdep}, we present the evolutionary tracks for a $5\;\msun$ star with $[{\rm Fe/H}] =0.0$ by varying the 
$\alpha_{\Lambda}$ parameter. This parameter has very little impact on the extent and luminosity of the blue loops. It mainly increases the RGB temperature as we increase the value of $\alpha_{\Lambda}$. This is similar to effect of the mixing-length parameter of MLT models. Next, we show the impact of $\alpha_{\omega}$ on the evolutionary track of a $5\;\msun$ star with $[{\rm Fe/H}] =0.0$ in the lower panels of Fig.~\ref{fig:paramdep}. This parameter determines the extent of the overshooting region as it indicates the impact of the non-local flux of the TKE \citep{ahlb22}. Hence, this parameter has a strong influence on the extent and luminosity level of the blue loops (lower left panel). Larger $\alpha_{\omega}$ leads to a larger core and an overall higher luminosity. It also decreases the extent of the blue loop. This behavior is analogous to the effects of the overshooting parameter in MLT models. The extent of the blue loop can be primarily characterised by the core potential (\(\approx\text{core mass}/{\text{core radius}}\)) \citep{kipp13}. A smaller value of this ratio results in a more extended blue loop and vice versa. Overshooting on the main sequence increases this ratio with increasing non-local parameter \( \alpha_{\omega} \)  or overshooting parameter $f_{\rm OV}$ and, consequently, results in a less extended blue loop. Fixing $\alpha_{\omega}$ during the MS phase and varying it only in the post-MS phase has a minimal impact on the blue loops (lower right panel). Thus, non-locality on the MS has the largest impact on the emergence of a blue loop. A detailed investigation of the influence of each of these parameters in the various evolutionary phases goes beyond the purpose of this study, and therefore only metallicity, as an observational parameter, was varied, when we compared our models to observed \cephs.

Out of the five systems tested, our models successfully reproduced the observations for OGLE-LMC-CEP-0227 and OGLE-LMC-CEP-2532. The fact that these two systems have the most precisely estimated parameters highlights the importance of observational precision for a better understanding and tighter constraints on the underlying evolutionary physics. For the other three systems, we find similar problems in obtaining the correct values from the models as was reported for modeling efforts done with MLT plus overshooting.

The \ceph\  of the OGLE-LMC-CEP-1812 system is predicted to be in a very short-lived stellar evolutionary phase, which is not impossible but rather unlikely. Additionally, the less massive, non-pulsating companion is found to be in a more advanced evolutionary stage. This may point to a complex binary evolutionary scenario, potentially involving mass transfer, binary mergers, or other interactions, a detailed investigation of which is beyond the scope of this paper \citep{neil15}. In this case, our treatment of two single stars evolving independently would not be applicable.

In the case of OGLE-LMC-CEP-4506, there is a possibility of a large uncertainty in estimating the effective temperature from observations due to its relatively long orbital period. Currently, the MLT plus overshooting model appears to fit this system better than TCM1, as it predicts a slightly higher luminosity. However, this conclusion is also uncertain, given the lack of precise metallicity estimates for any of the binary systems considered in this study, as we discuss further below.

Our models neither can fit simultaneously the observations for both components of OGLE-LMC-CEP-1718. Given that both stars in this system are pulsating, the complexities in their light curve modeling may have introduced uncertainty in the estimated parameters (also see \citealt{pile18}). 
A possible alternative to fitting both stars with the same metallicity
($\sim -0.3$ as in Fig.~\ref{fig:ogle1718_0.3}) involves widening the gap between the second and third crossings in terms of luminosity. This could be achieved with a suitable choice of free parameters in TCM1. As shown in Fig.~\ref{fig:blueloop_comp}, considering non-locality in different regions of the stars alters this gap, suggesting that tuning the parameters might offer a solution.
In the future, a thorough parameter study of the free parameters of the TCM1 could help in improving the models to better match with observations and also find the ideal parameter set to use for stellar models.

Another important future aspect of this study will be estimating a mass-luminosity relation for \cephs\ based on evolutionary models incorporating TCM1. This mass-luminosity relation is crucial for constructing pulsation model grids. In previous studies, the mass-luminosity relation is based on MLT plus overshooting evolutionary models \citep{bono00,ande14} which is a time-independent convection theory and the overshooting is an ad hoc, empirical fix for the shortcomings of the local MLT. 
However, for pulsation calculations, we need a non-local and time-dependent convection treatment. For example, \textsc{mesa-rsp} uses a modified version of the \citet{kuhf86} model. As a result of these differing convection treatments, it remains challenging to directly continue pulsational model computation from the evolutionary models' envelope \citep{paxt19}. Moreover, the light curve structures of radially pulsating stars are highly sensitive to the convection treatment in their envelopes. In our future work, we aim to explore the integration of TCM3 into both the evolutionary and pulsation models of radially pulsating stars. We anticipate that this approach could improve our ability to model the observed properties of these stars with greater accuracy, though its effectiveness will depend on how well the TCM3  captures the underlying physical processes.

\section{Conclusions}
\label{sec:conclusion}
The evolutionary tracks of \cephs, especially the appearance of the blues loops, are highly sensitive to the input physics of the stellar models. In combination with the highly accurate observations of binary systems, this makes \cephs\ ideal laboratories to test new physics of stellar evolutionary models. In this study, we used the observations of \cephs\ in eclipsing binary systems to test the predictions of the TCM1 convection model \citep{kuhf86,kuhf87}. Furthermore, we investigated the influence of CBM on the structure and evolution models of \cephs, by testing how the blue loop changes if we apply CBM only at specific boundaries, followed by a parameter study of the most important free parameters of the TCM1.

We show above that the solution of the TCM1 equation without any external descriptions or any fine-tuning of the numerical parameters involved predicts the overshooting zone beyond the Schwarzschild boundary. This leads to the emergence of blue loops in the evolutionary tracks of intermediate mass stars, which is a necessity to explain the observations of \cephs. 

We do find that the inclusion or neglect of non-locality at different boundaries in the post-MS evolution slightly modifies the morphology and extent of the blue loops (see Fig~\ref{fig:blueloop_comp}). However, the non-locality in the MS phase seems to be most decisive for the emergence of a blue loop. Furthermore, the model effectively constrains the temperature gradient in the overshooting zone which the ad-hoc overshooting in MLT implementation can not predict. However, as shown by  \cite{tere24}, TCM1 struggles to accurately model the temperature stratification in the solar convective envelope. The application of TCM3 is expected to model convection more accurately in the envelopes of 1D stellar evolutionary models \citep{tere24}.

The ``\ceph\  mass discrepancy'' has long been a challenging puzzle in astrophysics. One proposed solution that has been put forward to solve this discrepancy concerns the size of the MS convective core (Sect.~\ref{sec:3.2}). Previously, this had been addressed by incorporating overshooting in an ad-hoc manner \citep[e.g.,][]{cass11,moro12}. This mainly involves adjusting the overshooting parameters to match observations without considering the underlying physics of convection. The comparison of TCM1 models to the observations of five \ceph\ systems shows that TCM1 reproduces the observations with similar accuracy as MLT models including ad hoc overshooting. This is achieved without fine-tuning any of the TMC1 model parameters. 

From comparing the evolutionary tracks in Figs.~\ref{fig:ogle0227_0.5} -- \ref{fig:ogle1718_0.5} it is apparent that the TCM1 blue loops are slightly more extended, and that
the luminosity of the TCM1 models is generally lower than for the MLT plus overshooting tracks. This difference increases slightly with decreasing mass, even if the mass range of the \cephs\ investigated here is rather limited.
These luminosity differences could probably be minimized or even eliminated by fine-tuning the $\alpha_{\omega}$ parameter. However, our primary focus is on reproducing the observations using the same values of the free parameters involved in TCM1, rather than adjusting them individually for each observation. Hence, we conclude that TCM1 predicts the overshooting at convective boundaries very accurately for the default parameters.

Finally, TCM1 still relies on certain assumptions about the free parameters, which are calibrated using the local and time-independent MLT plus overshooting model \citep[e.g.][]{ahlb22b}. These assumptions are not mandatory. A more effective approach to constrain these parameters would be to integrate evolutionary models with full amplitude stable mode pulsation models and compare their predictions with observations, particularly for radially pulsating stars. Additionally, carrying out a comparison as done in this work with the more physically complete TCM3 would be the next step towards better modeling of convection in stellar evolution codes.

\section*{Acknowledgements}
 We thank the reviewer for the valuable and constructive feedback and suggestions. We express our gratitude to the Kavli Foundation and the Max Planck Institute for Astrophysics for their support of the Kavli Summer Program in Astrophysics, 2023. This program provided the environment in which a significant portion of this project was undertaken. MD thanks Marcella Marconi (Istituto Nazionale di Astrofisica, INAF – Osservatorio Astronomico di Capodimonte) for her valuable feedback and acknowledges funding from the INAF 2023 Large Grant MOVIE (PI: Marcella Marconi). FA acknowledges funding from the ERC Consolidator Grant DipolarSound (grant agreement \# 101000296) and support from the Klaus Tschira foundation.

\bibliographystyle{aa}

\appendix
\clearpage

\onecolumn

\section{Chemical composition}
\label{sec:feh}
All the [Fe/H] values and the corresponding $X$, $Y$ and $Z$ adopted in this work are given in Table~\ref{table:feh}.

\begin{table}[h]
	\centering
	\caption{Chemical compositions of the adopted models.}
	\label{table:feh}
	\begin{tabular}{lccccccr} 
		\hline
        [Fe/H] & $X$&$Y$&$Z$ \\ \hline
        $\hphantom{-}0.0$ & $0.7152$ & $0.2665$ & $0.0183$ \\
        $-0.6$ & $0.7429$ & $0.2527$ & $0.0043$ \\
        $-0.5$ & $0.7407$ & $0.2538$ & $0.0054$ \\
        $-0.3$ & $0.7346$ & $0.2568$ & $0.0085$ \\
        $\hphantom{-}0.3$ & $0.6886$ & $0.2796$ & $0.0317$ \\
        $\hphantom{-}0.6$ & $0.6357$ & $0.3057$ & $0.0584$ \\         
      \hline
	\end{tabular}
\end{table}

\section{Stellar parameters for the binary systems considered in this work} \label{sec:binpar}

The stellar parameters of the binary systems obtained from the literature are given in Table~\ref{table:tab2}. We have chosen five of the systems described by \citet{pile13} and \citet{pile18} to 
test the TCM1 and have included  them in the table. 

 \begin{table*}[h]
        \centering
        \caption{Properties of the eclipsing binary systems obtained from the literature. ``F'' and ``FO'' denote the fundamental mode and first-overtone mode of \cephs\ , respectively.}
        \label{table:tab2}
        \begin{tabular}{lcccccccccr} 
        \hline
        OGLE ID& Parameter & Primary & Secondary  \\   \hline
	      OGLE-LMC-CEP-0227      &Orbital period (days)  &\multicolumn{2}{c}{309.404}  \\
             (F)              &Pulsation period (days)    & $3.797$& \ldots  \\
        Source: \citet{pile13}                   &$M/\msun$             & $4.165\pm0.032$&$4.134\pm0.037$ \\
                           &$R/\rsun$              & $34.92\pm0.29$&$44.85\pm0.34$ \\
                           &$\log{(L/\lsun)}$              & $3.158\pm0.049$&$3.097\pm0.047$ \\
                           &$T_{\rm eff}$(K)              & $6050\pm160$&$5120\pm130$ \\ \hline
        OGLE-LMC-CEP-1812      &Orbital period (days)  &\multicolumn{2}{c}{551.8}  \\
             (F)              &Pulsation period (days)    & $1.313$& \ldots \\ 
          Source: \citet{pile18}                 &$M/\msun$             & $3.76\pm0.03$&$2.62\pm0.02$ \\
                           &$R/\rsun$              & $17.85\pm0.13$&$11.83\pm0.08$ \\
                           &$\log{(L/\lsun)}$              & $2.61\pm0.04$&$1.95\pm0.04$ \\
                           &$T_{\rm eff}$(K)              & $6120\pm150$&$5170\pm120$ \\ \hline
	    OGLE-LMC-CEP-4506      &Orbital period (days)  &\multicolumn{2}{c}{1550}  \\
             (F)              &Pulsation period (days)    & $2.988$& \ldots\\
         Source: \citet{pile18}                   &$M/\msun$             & $3.61\pm0.03$&$3.52\pm0.03$ \\
                           &$R/\rsun$              & $28.5\pm0.2$&$26.4\pm0.2$ \\
                           &$\log{(L/\lsun)}$              & $3.01\pm0.05$&$2.93\pm0.05$ \\
                           &$T_{\rm eff}$(K)              & $6120\pm160$&$6070\pm150$ \\ \hline
	    OGLE-LMC-CEP-1718     &Orbital period (days)  &\multicolumn{2}{c}{412.8} \\
            (FO+FO)               &Pulsation period (days)    & $1.964$&$ 2.481$ \\
         Source: \citet{pile18}                   &$M/\msun$             & $4.27\pm0.04$&$4.22\pm0.04$ \\
                           &$R/\rsun$              & $27.8\pm1.2$&$33.1\pm1.3$ \\
                           &$\log{(L/\lsun)}$              & $3.04\pm0.06$&$3.18\pm0.06$ \\
                           &$T_{\rm eff}$(K)              & $6310\pm150$&$6270\pm160$ \\ \hline
	    OGLE-LMC-CEP-2532      &Orbital period (days)  &\multicolumn{2}{c}{800.4} \\
           (FO)                &Pulsation period (days)    & $2.035$& \ldots \\
         Source: \citet{pile18}                   &$M/\msun$             & $3.98\pm0.10$&$3.94\pm0.09$ \\
                           &$R/\rsun$              & $29.2\pm1.4$&$38.1\pm1.8$ \\
                           &$\log{(L/\lsun)}$              & $3.10\pm0.06$&$2.84\pm0.09$ \\
                           &$T_{\rm eff}$(K)              & $6350\pm150$&$4800\pm220$ \\ \hline

        \end{tabular}
\end{table*}

\section{Evolutionary tracks for OGLE-LMC-CEP-0227 with inverted mass ratio}\label{sec:invmass0227}

\begin{figure*}[hb!]
\includegraphics[width=0.9\textwidth,keepaspectratio]{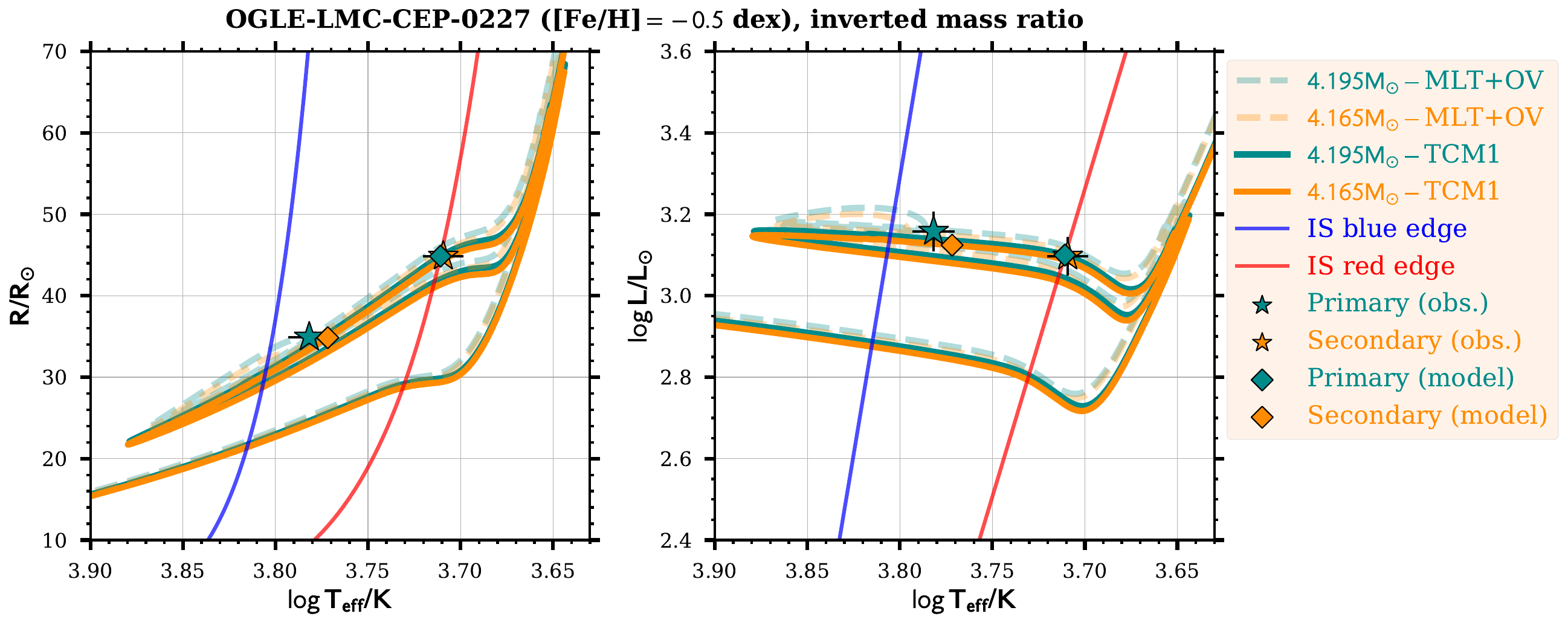}
\caption{The alternate models for OGLE-LMC-CEP-0227 by assigning inverted mass to primary and secondary components. This figure is discussed in detail in Sect.~\ref{sec:ogle227}.}
\label{fig:ogle0227_0.5_alt}
\end{figure*}

Fig.~\ref{fig:ogle0227_0.5_alt} presents the evolutionary tracks for OGLE-LMC-CEP-0227, plotted with inverted mass ratio compared to the observations, as discussed in Sect.~\ref{sec:ogle227}.

\section{Evolutionary tracks with different metallicities}
\begin{figure*}[htb!]
\includegraphics[width=1.0\textwidth,keepaspectratio]{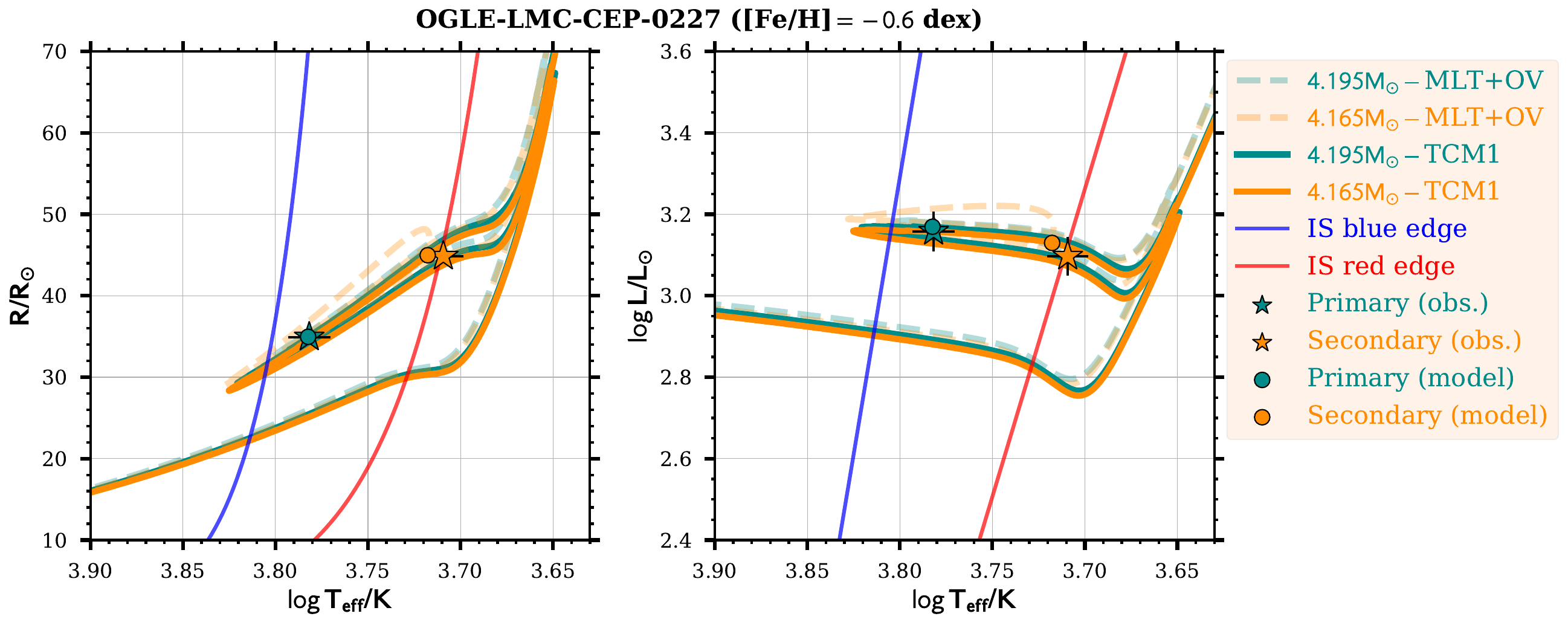}
\caption{As Fig.~\ref{fig:ogle0227_0.5} but with different [Fe/H] value ($-0.6$). Some error bars are smaller than the symbols. This figure is discussed in detail in Sect.~\ref{sec:ogle227}.}
\label{fig:ogle0227_0.6}
\end{figure*}

\begin{figure*}[htb!]
\includegraphics[width=1.0\textwidth,keepaspectratio]{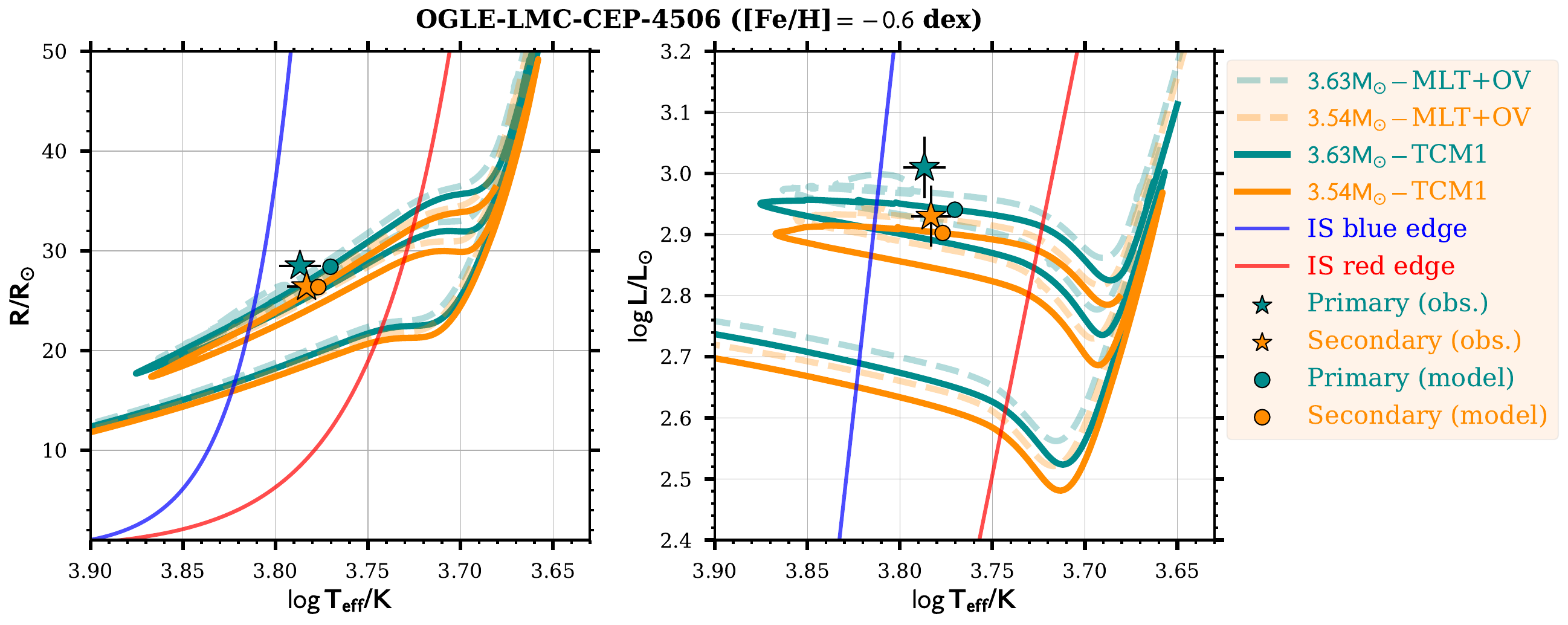}
\caption{Same as Fig.~\ref{fig:ogle4506_0.5} but with a different metallicity,
$\rm{[Fe/H]=-0.6}$. This figure is discussed in detail in Sect.~\ref{sec:ogle4506}.}
\label{fig:ogle4506_0.6}
\end{figure*}

\begin{figure*}[htb!]
\includegraphics[width=1.0\textwidth,keepaspectratio]{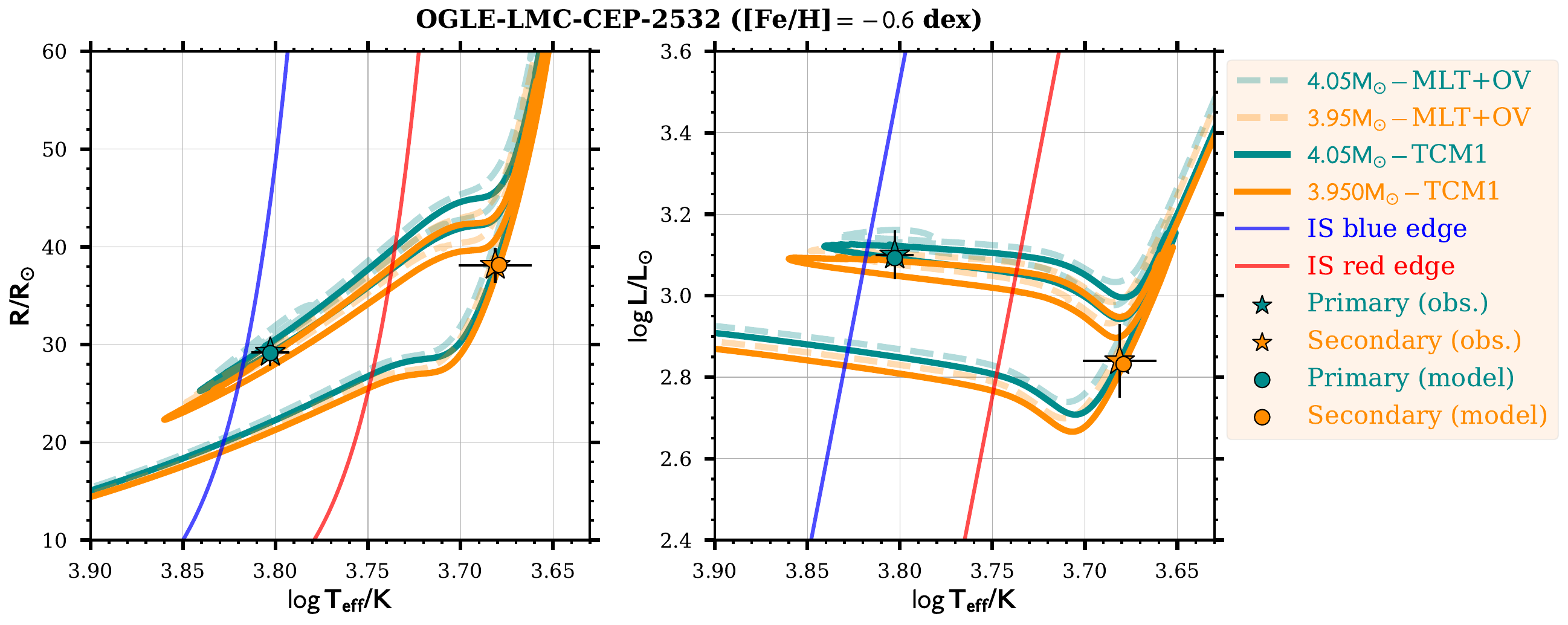}
\caption{Same as Fig.~\ref{fig:ogle2532_0.5} but with a different metallicity,
$\rm{[Fe/H]=-0.6}$. This figure is discussed in detail in Sect.~\ref{sec:ogle2532}.}
\label{fig:ogle2532_0.6}
\end{figure*}

\begin{figure*}[htb!]
\includegraphics[width=1.0\textwidth,keepaspectratio]{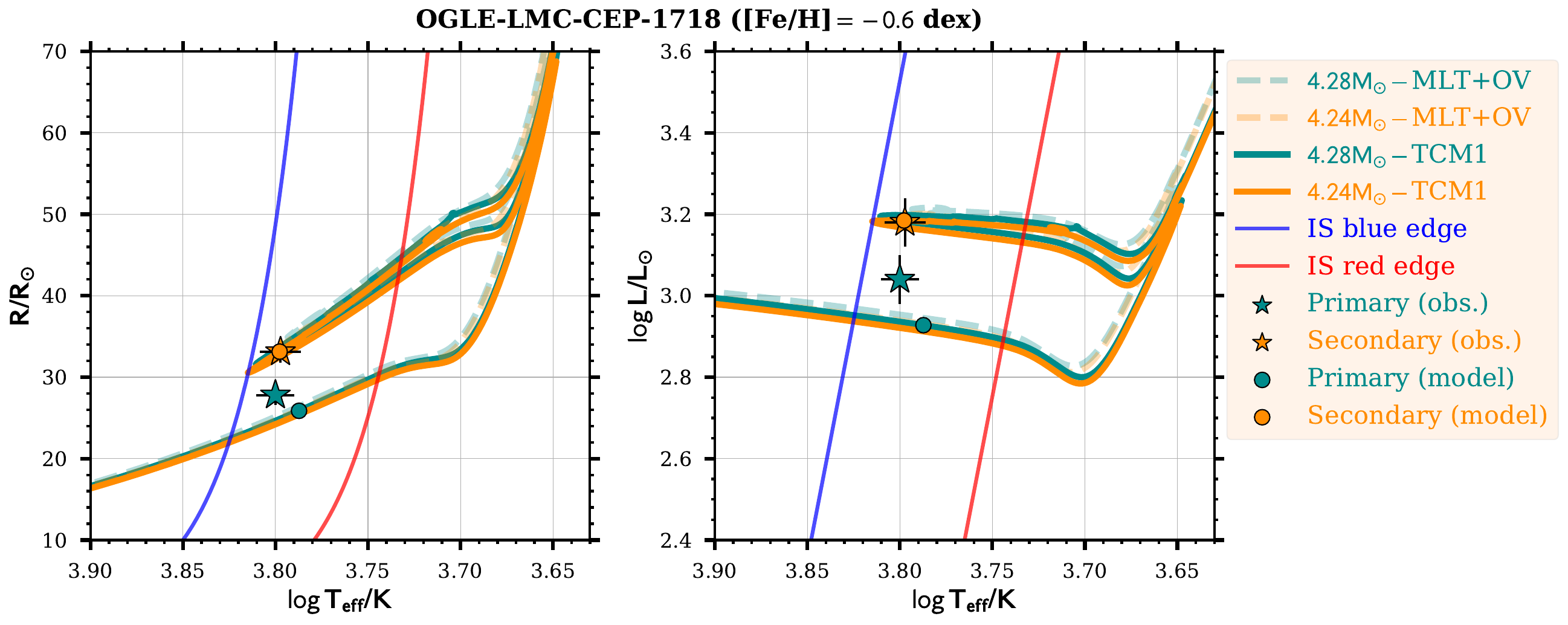}
\caption{Same as Fig.~\ref{fig:ogle1718_0.5} but with a different metallicity,
$\rm{[Fe/H]=-0.6}$. This figure is discussed in detail in Sect.~\ref{sec:ogle1718}.}
\label{fig:ogle1718_0.6}
\end{figure*}

\begin{figure*}[htb!]
\includegraphics[width=1.0\textwidth,keepaspectratio]{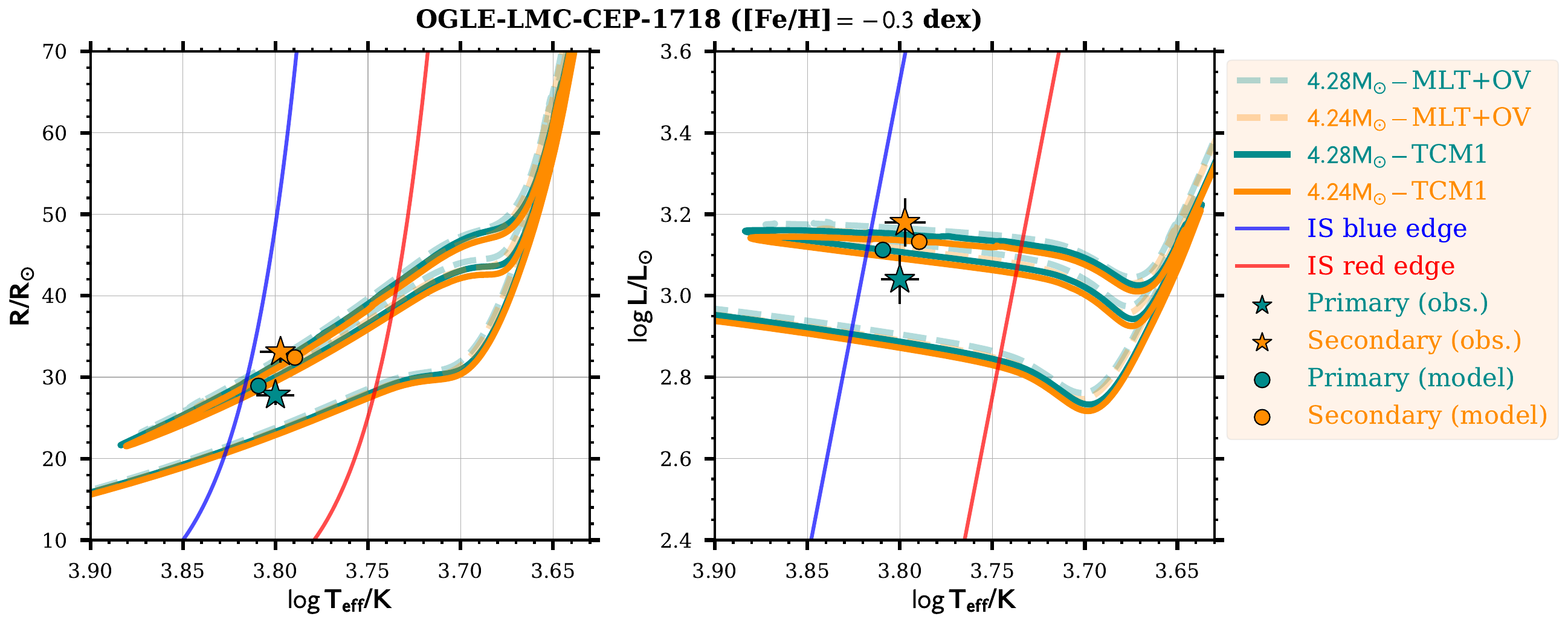}
\caption{Same as Fig.~\ref{fig:ogle1718_0.5} but with a different metallicity,
$\rm{[Fe/H]=-0.3}$.  This figure is discussed in detail in Sect.~\ref{sec:ogle1718}.}
\label{fig:ogle1718_0.3}
\end{figure*}

This section provides supplementary figures (Figs.~\ref{fig:ogle0227_0.6} -- \ref{fig:ogle1718_0.3}) of the evolutionary tracks of the observed binary systems with different metallicities as discussed in Sect.~\ref{sec:3.3}.

\clearpage

\end{document}